\documentclass{article}
\usepackage[utf8]{inputenc}
\usepackage[T1]{fontenc}
\usepackage{microtype}
\usepackage{graphicx}
\usepackage{booktabs}
\usepackage{tabularx}
\usepackage{subcaption}
\AtBeginDocument{\raggedbottom}
\usepackage{amsmath,amssymb}
\usepackage{enumitem}
\usepackage{pifont}
\usepackage{xcolor}
\usepackage{hyperref}
\usepackage{xurl}

\usepackage{mlsys2025}

\newcolumntype{Y}{>{\raggedright\arraybackslash}X}
\makeatletter
\newcommand{\pagenum}{\raisebox{-0.45in}[0pt][0pt]{\small\thepage}}
\AtBeginDocument{\fancyfoot[C]{\pagenum}}
\fancypagestyle{firstpage}{\fancyhf{}\renewcommand{\headrulewidth}{0pt}\fancyfoot[C]{\pagenum}}
\makeatother

\mlsystitlerunning{\mbox{DMuon-2}: Scaling Matrix Optimizers in Data Parallelism with Topology-Aware Matrix Ownership and Communication}

\AtBeginDocument{\hypersetup{pdfauthor={Starrick Liu, Ruizhi Chengze, Lufang Chen, Michael Zhao, Ran Yu, Xiaofan Li, dance kuzi, Yue Zhang, Zijuan Lin, SongAo Han, YongHua Chen, Ruyi Gan, Hao Wang}}}
\makeatletter
\renewcommand{\printAffiliationsAndNotice}[1]{%
\begingroup\renewcommand{\thefootnote}{*}%
\footnotetext{Corresponding authors: Starrick Liu (\href{mailto:starrick@x2robot.com}{\textcolor{black}{\textnormal{starrick@x2robot.com}}}); Hao Wang (\href{mailto:wanghao@x2robot.com}{\textcolor{black}{\textnormal{wanghao@x2robot.com}}}).}%
\endgroup}
\makeatother

\renewcommand{\icmlruler}[1]{}
\makeatletter
\renewcommand{\Notice@String}{}
\makeatother
\AtBeginDocument{\hypersetup{pdfsubject={Research preprint}}}

\begin{document}

\twocolumn[
\mlsystitle{\mbox{DMuon-2}: Scaling Matrix Optimizers in Data Parallelism with Topology-Aware Matrix Ownership and Communication}

\begin{center}\begin{minipage}{0.96\textwidth}\centering
{\normalsize\bfseries \mbox{Starrick Liu\textsuperscript{*}}, \mbox{Ruizhi Chengze}, \mbox{Lufang Chen}, \mbox{Michael Zhao}, \mbox{Ran Yu}, \mbox{Xiaofan Li}, \mbox{dance kuzi}, \mbox{Yue Zhang}, \mbox{Zijuan Lin}, \mbox{SongAo Han}, \mbox{YongHua Chen}, \mbox{Ruyi Gan}, \mbox{Hao Wang\textsuperscript{*}}\par}
\vspace{5pt}{\normalsize X Square Robot\par}
\end{minipage}\end{center}
\mlsyskeywords{Muon, FSDP, HSDP, Collective Communication}

\vskip 0.3in

\begin{abstract}
Matrix optimizers such as Muon update whole weight matrices, which conflicts with the element- or row-wise sharding of data-parallel runtimes. Once parameters are sharded as well, matrix optimizers incur extra communication or repeated computation. We present \mbox{DMuon-2}, a matrix-granular optimizer runtime for PyTorch FSDP (ZeRO-2/3) and HSDP. \mbox{DMuon-2} assigns each matrix a unique owner, which performs the Newton--Schulz (NS) iteration and optimizer update on the full matrix, and balances the load across owners by computational cost. On HSDP, ownership is global across replicas, so the cluster keeps a single copy of gradients and optimizer state, and updated parameters are published across nodes as soon as they are ready, overlapping subsequent computation. This layout makes every collective sparse and variable in length, so \mbox{DMuon-2} schedules these collectives as their data becomes ready and routes them along the cluster's two-level network. On up to 256 H100 GPUs, across three dense LLMs and two VLA models, \mbox{DMuon-2} matches or exceeds the training throughput of PyTorch FSDP2 with AdamW, reaching 1.05$\times$ that of FSDP2 with AdamW at the median on 256 GPUs. The optimized parameter gathering and gradient reduction absorb the extra communication and computation a matrix optimizer requires. Against a gather-then-replicate Muon baseline, \mbox{DMuon-2}'s throughput is 2.1 times as high at the median.
\end{abstract}
]

\printAffiliationsAndNotice{}
\thispagestyle{firstpage}

\section{Introduction}

Muon~\citep{jordan2024muon} brings orthogonalization into neural-network optimization. Unlike AdamW~\citep{loshchilov2019adamw}, the object Muon updates is a whole two-dimensional weight matrix, and cutting a matrix into shards and running Newton--Schulz (NS) iteration on each one is in general not equivalent to running NS once on the whole matrix. Training at frontier scale, meanwhile, relies on Fully Sharded Data Parallel (FSDP, i.e., ZeRO-2/3) and its Hybrid Sharded Data Parallel (HSDP) variant to spread model state~\citep{zhao2023fsdp}, which rests on the assumption that element-wise updates commute with sharding. Matrix optimizers break this assumption.

Existing work handles this conflict along two lines. The first gathers the gradient shards of all ranks back into whole matrices before the optimizer step and runs NS on every rank, which preserves the semantics at the cost of an extra gather and repeated NS computation~\citep{liu2025muon}. The second makes a few ranks responsible for each matrix and, on ZeRO-1, already removes the extra communication from the optimizer step~\citep{lim2025motif,wang2026canzona,stepfun2026step35,deepseek2026v4}, but in all of these approaches the parameters are fully replicated on every rank. On parameter-sharded runtimes such as FSDP and HSDP, no work yet removes both the repeated computation and the extra communication at once (Section 8).

\mbox{DMuon-2} makes exactly one rank responsible for each matrix, called its owner. In the backward pass, the matrix's gradient is reduced from all ranks to the owner, the owner performs one NS and one optimizer update, and in the forward pass the new parameters are broadcast back to all ranks. These two communications replace exactly the ReduceScatter and AllGather that FSDP already performs, without adding a communication phase. HSDP confines sharding to the node, yet each replica still runs the same optimizer update, which for Muon means the same NS repeated on every replica. \mbox{DMuon-2} keeps this pattern, but each matrix has a single owner in the whole cluster, the gradient is reduced to it, the optimizer state is stored only on it, and NS runs only once.

The shape of the communication changes accordingly. The matrices of a layer belong to different owners and differ in size, so each communication is rooted at only a few owners and its segments have different lengths. It is a sparse, variable-length ReduceScatter or AllGather. Existing communication libraries support this pattern poorly, and primitives for the reduction side are particularly lacking~\citep{wang2026vescale}. These communications also cross two network levels with very different bandwidths. \mbox{DMuon-2} follows the established idea of organizing collective communication by network level~\citep{shah2023taccl,hidayetoglu2024hiccl,cao2025syccl} and implements three sparse, variable-length operators.

\mbox{DMuon-2} makes three contributions.
\begin{enumerate}[leftmargin=*,topsep=2pt,itemsep=1pt,parsep=0pt]
\item \textbf{Communication-free matrix ownership.} We define owners at matrix granularity, uniformly for FSDP (ZeRO-2/3) and HSDP. Each matrix runs NS and its optimizer update once, on its owner, at the same logical communication volume as FSDP, and owners are assigned by computational cost so that NS work stays balanced as the cluster grows.
\item \textbf{Globally unique ownership across HSDP replicas.} On HSDP, owners are unique across replicas, so the cluster keeps a single copy of gradients and optimizer state. A cross-step publish returns each updated parameter to every node as soon as it is ready, overlapping subsequent computation.
\item \textbf{Topology-aware variable-length communication operators.} We implement AllGatherV, ReduceScatterV, and the cross-node publish as single kernels that send or forward data as soon as it is ready instead of in a fixed order, complete multicast and reduction in one hop inside a node, and shrink the chain from the number of ranks to the number of nodes.
\end{enumerate}
We evaluate \mbox{DMuon-2} on up to 32 nodes with 256 H100 GPUs, under several FSDP and HSDP sharding configurations, on three dense LLMs and two VLA models. \mbox{DMuon-2} matches or exceeds the training throughput of PyTorch FSDP2 with AdamW, reaching 1.05$\times$ its throughput at the median on 256 GPUs, and is 2.1 times as fast at the median as Muon implemented along the first route. \mbox{DMuon-2} is open source at \url{https://github.com/X-Square-Robot/dmuon} and has been used in the pre-training of two production foundation models, Wall-OSS-0.5~\citep{yu2026walloss} and WALL-WM~\citep{li2026wallwm}.

\section{Background}

\subsection{Muon and Matrix-Granular Updates}

Matrix optimizers treat the whole weight matrix as the basic unit of update. Examples include Distributed Shampoo~\citep{shi2023shampoo}, Muon~\citep{jordan2024muon}, and Dion~\citep{ahn2025dion}. Among them, Muon reaches a given loss with less compute than AdamW in large-scale pre-training and has been used to pre-train models with hundreds of billions to trillions of parameters~\citep{liu2025muon,kimi2026k3,deepseek2026v4}.

Muon applies only to two-dimensional weight matrices. Embeddings, the output head, and the gains and biases of normalization layers are still updated by element-wise optimizers such as AdamW~\citep{jordan2024muon}. For $W\in\mathbb{R}^{m\times n}$, let $G(W)$ denote the average of the local gradients of the $N$ data-parallel ranks in the current step. Muon first updates the momentum with it, then orthogonalizes the momentum and updates the weight,
\begin{equation*}
M_t=\mu M_{t-1}+G(W),\qquad W_{t+1}=W_t-\eta\,\mathrm{NS}(M_t),
\end{equation*}
where $\mu$ is the momentum coefficient and $\eta$ the learning rate; our implementation uses the Nesterov variant of this momentum. The momentum has the same shape as $W$, is kept in FP32, and is Muon's entire optimizer state, half of AdamW's.

NS denotes Newton--Schulz iteration, which approximates the orthogonal factor of the polar decomposition of $M_t$, the semi-orthogonal matrix closest to it, with a few polynomial iterations~\citep{jordan2024muon}. Every iteration is a matrix multiplication on the whole matrix. With $m\le n$, its cost is proportional to $m^2n$.

Orthogonalization cannot be split across shards. Running NS on each shard separately agrees with running NS once on the whole matrix if and only if the rows of different shards are pairwise orthogonal, as derived in Appendix A. The rows of a linear layer's gradient are all linear combinations of the same batch of input activations and are therefore correlated, as is the momentum, a weighted sum of gradients, so the condition generally fails. Muon must therefore update whole matrices, and in sharded training every matrix must first be assembled on some rank.

\subsection{FSDP and HSDP}

Large-model training must distribute model state over many GPUs. The standard approach is the sharding strategy of ZeRO~\citep{rajbhandari2020zero}, of which PyTorch's FSDP~\citep{zhao2023fsdp} is an implementation.

PyTorch's FSDP2 splits every parameter evenly across $N$ ranks and communicates layer by layer, gathering a layer's parameters with one AllGather before computing it and reducing its gradients onto the shards with one ReduceScatter after the backward pass. After the forward pass, FSDP either frees the full parameters and gathers them again before the backward pass (ZeRO-3), or keeps them until the end of the backward pass at the cost of more memory (ZeRO-2).

In multi-node training, every per-layer communication of FSDP crosses the inter-node network. HSDP~\citep{zhao2023fsdp,zhang2022mics} therefore arranges the $N$ ranks into an $R\times K$ mesh with $N=RK$, where $R$ usually equals the number of nodes and $K$ the number of GPUs per node, and denotes rank coordinates by $(r,s)$. Each replicate row is usually one node and shards within itself as in FSDP. The $R$ ranks sharing shard coordinate $s$ form a replicate group and hold the same shard. Per-layer AllGather and ReduceScatter run only within the row, after which an AllReduce over the gradient shards runs within each replicate group. This synchronization is the only cross-node traffic, and the optimizer step runs once in every replicate row.

With bandwidth-optimal implementations, each rank receives and sends $\frac{N-1}{N}P$ bytes in one AllGather or ReduceScatter, where $P$ is the total number of bytes involved. Each rank therefore communicates $2\frac{N-1}{N}P$ bytes per step under ZeRO-2 and $3\frac{N-1}{N}P$ under ZeRO-3, and HSDP keeps this per-layer communication inside the node, leaving one gradient synchronization across nodes.

This sharding assumes an element-wise optimizer, with each rank updating only its own shard. Muon needs whole matrices, so exact updates on FSDP require extra communication in the optimizer step. Distributed Muon~\citep{liu2025muon} gathers every matrix on all ranks and runs NS on each, so every rank receives an extra $\frac{N-1}{N}P_M$ bytes and NS is repeated $N$ times, where $P_M$ is the total size in bytes of the matrices Muon manages. On HSDP, even if each replicate row runs NS only once, the whole cluster still repeats it $R$ times and stores the optimizer state $R$ times.

\subsection{Two-Level Networks and Collective Communication in GPU Clusters}

Typical GPU training clusters use a two-level network, with NVLink inside the node and an RDMA network between nodes. In an H100 server, the 8 GPUs of a node are fully connected through NVSwitch, and third-generation NVSwitch can also perform multicast and reduction inside the switch, known as NVLink SHARP, or NVLS, which is accessed with the multimem family of PTX instructions~\citep{andersch2022hopper,nvidia2026cuda}.

Between nodes, each GPU usually has one InfiniBand or RoCE NIC. The $i$-th GPU and the $i$-th NIC form the $i$-th rail, and common fabrics connect the same-numbered NICs of all nodes to the same group of switches~\citep{nvidia2026superpod}, with a single rail carrying nearly an order of magnitude less bandwidth than NVLink. For a GPU to send over another rail, it must first hand the data over NVLink to another GPU in the same node. The replicate groups of HSDP usually consist exactly of the same-numbered GPUs of all nodes, so their communication uses a single rail.

Cross-node communication has traditionally been driven by proxy threads on the CPU. Recent NCCL versions provide a device-side API through which a kernel can read and write the memory of other GPUs in the same node over NVLink, and can write data to other nodes and update the peer's signal through GPU-initiated RDMA (GPU-Initiated Networking, GIN), without involving the CPU~\citep{hamidouche2025gin}, so a single kernel can perform transfers at both levels.

NCCL's AllGather and ReduceScatter require equal-length data on all ranks, its public API has no variable-length versions, and although it implements AllGatherV internally, the reduction side has no counterpart~\citep{nvidia2026nccl}. PyTorch's single-tensor interfaces also require equal lengths, and its list interfaces fall back to one Broadcast or Reduce per rank when lengths differ~\citep{pytorch2026pgnccl}.

Collective communication is itself a GPU kernel. Each NCCL channel corresponds to one thread block and occupies one SM, and the number of channels grows with the message size. When communication overlaps computation, the SMs occupied by the communication kernel cannot be used for computation during that time, so the more it takes, the less is left for computation.

\begin{figure*}[t]
\centering
\includegraphics[width=\textwidth]{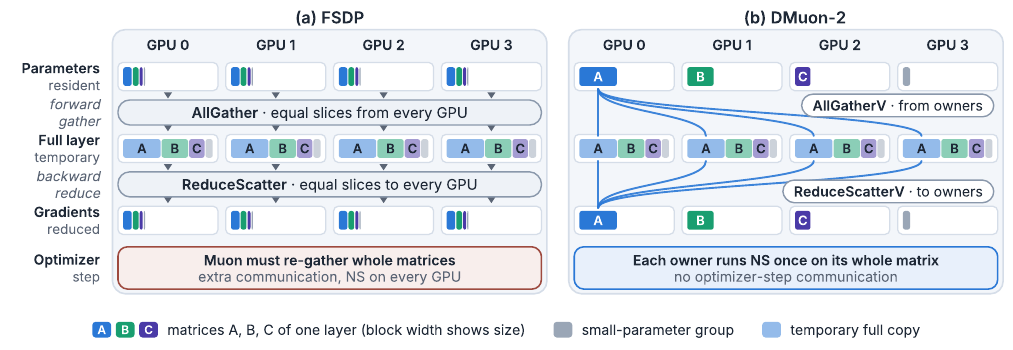}
\caption{Communication of one layer under FSDP and \mbox{DMuon-2}, with four GPUs and a layer containing matrices A, B, and C and a group of smaller parameters; in (b) every matrix belongs in full to one owner. Only the paths of matrix A are drawn. B, C, and the small-parameter group are the same.\label{fig:1}}
\end{figure*}

\section{Matrix Owners on FSDP}

\subsection{Matrix-Granular Sharding}

\mbox{DMuon-2} raises the sharding granularity from row blocks to whole matrices, giving each matrix to exactly one rank, its owner. The owner keeps the matrix's full parameters, momentum, and reduced gradient, and NS and the optimizer update run only on the owner, once. Other ranks hold no shard of the matrix and obtain its full parameters temporarily only when the forward or backward pass needs them. Parameters, gradients, and optimizer state are still each stored once in the whole cluster, as in FSDP. For such an update to be equivalent to single-device training, the owner must obtain the average of all ranks' gradients before updating, and every other rank must read the updated parameters before using the matrix again. Reduction and broadcast guarantee these two conditions respectively.

\subsection{Owner-Rooted Communication and Its Volume}

These two conditions determine how FSDP's two kinds of communication change (Figure~\ref{fig:1}). Before a layer is computed in the forward pass, the matrix's full parameters exist only on the owner, so FSDP's AllGather becomes a Broadcast rooted at the owner, as does the AllGather before the backward pass under ZeRO-3. After the layer's gradient is computed in the backward pass, the gradient must be reduced in full to the owner, so ReduceScatter becomes a Reduce rooted at the owner. \mbox{DMuon-2} keeps FSDP's per-layer communication, including its prefetching and overlap with computation. The multiple Broadcasts and Reduces of a layer together form one AllGatherV and one ReduceScatterV. Because the owner already receives the complete gradient, the optimizer step itself needs no communication at all.

\mbox{DMuon-2} has the same logical communication volume as FSDP. For a matrix of $P_W$ bytes, each of the other $N-1$ ranks must send its own gradient to the owner and must receive all $P_W$ bytes back, so pipelined Reduce and Broadcast each move $(N-1)P_W$ bytes in total and cannot move less under this communication structure. Per rank and per step this gives $2\frac{N-1}{N}P_M$ under ZeRO-2 and $3\frac{N-1}{N}P_M$ under ZeRO-3, identical to FSDP's values in Section 2.2.

\subsection{Load Balancing}

Even sharding gives every rank the same computation. Assigning owners by matrix no longer does. Ranks hold whole matrices of different sizes, so after finishing their own NS they wait at the first communication of the next step for the busiest owner, which determines the duration of the optimizer step, and the state of every matrix lives on its owner. Both the NS time and the bytes held therefore have to be balanced across ranks. Assigning owners round-robin in parameter order leaves the busiest rank of the vision-language-action (VLA) model $\pi_{0.5}$~\citep{pi2025pi05} with 2.15 times the average NS time.

Owners are also communication roots, and balancing communication follows a different logic from balancing computation. Computation accumulates over all layers and only needs to be even over the whole model, whereas communication proceeds layer by layer and must also be even within a layer. When the matrices of one layer concentrate on a few ranks, their communications cannot proceed at the same time, the fill and drain cost is spread over fewer roots, and the inter-node bandwidth is left underused (Section 5).

\mbox{DMuon-2} works at the granularity of assignment units. One large matrix forms one unit, while the small matrices of a layer are merged into a single unit sharing one owner, so a layer never contains many tiny communications. A layer here coincides with an FSDP wrapping unit. The objective is to minimize the NS time of the busiest rank, subject to every rank holding at most $1+\epsilon$ times the average bytes, with $\epsilon=0.1$, and to the assignment units of one layer belonging to different ranks. The byte count of a unit is known in advance, but its NS time is not. It depends on matrix shape and is not proportional to the number of elements, so estimating it by element count or by $m^2n$ departs noticeably from measurement, and \mbox{DMuon-2} instead measures every matrix shape once before assignment. Under an element-count estimate, the busiest rank of Qwen2.5-1.5B on 32 ranks spends 1.81 times the mean NS time, no better than the 1.74 times it spends with no balancing at all (Appendix B). Even without the two constraints this is NP-hard multiprocessor scheduling, which \mbox{DMuon-2} approximates with the greedy Longest Processing Time (LPT) rule~\citep{graham1969bounds}, giving priority to layer spreading when the constraints conflict.

\begin{figure}[t]
\centering
\includegraphics[width=\columnwidth]{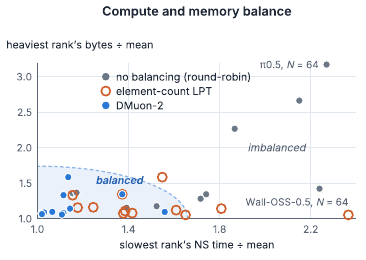}
\caption{Load balance of owner assignment. The four models other than Qwen2.5-32B assigned to 16, 32, and 64 ranks. The axes are the NS time of the busiest rank and the bytes held by the heaviest rank, both normalized to the per-rank average. The shaded region is the balanced region. The measurement protocol is given in Appendix B.\label{fig:2}}
\end{figure}

As the cluster grows, each GPU's share of NS shrinks with the GPU count, down to the time of the largest assignment unit (Figure~\ref{fig:2}).

\section{Global Owners on HSDP}

\subsection{Global Owners and Leaders}

The owner mechanism of Section 3, carried over to HSDP, can only spread NS over the GPUs of one node, and it does not shrink as nodes are added. \mbox{DMuon-2}'s layout on HSDP follows from two constraints. Muon requires the complete gradient sum of each matrix to meet in one place, and HSDP requires per-layer parameter gathering to stay inside the node. Satisfying both yields a global owner plus one leader per row. Plain HSDP updates the parameters once on every node because the cross-node AllReduce hands the summed gradient to every node, so \mbox{DMuon-2} splits this AllReduce into its two halves. The reduction half delivers the sum of matrix $W$'s gradients over all ranks of the cluster to a single rank, the global owner of $W$, which holds $W$'s momentum and FP32 master parameters and is the only rank to run NS and the update. No other rank holds that gradient sum or any of $W$'s optimizer state, so the update happens once in the whole cluster.

The return half becomes a cross-node publish of the parameters. Every node needs the full parameters of $W$ for its forward pass, so in each row \mbox{DMuon-2} has the GPU in the same column as the owner keep a full BF16 copy, the row's leader, and the owner writes the new parameters into these copies once it has updated. Before a layer is computed, the leader broadcasts this copy to the other GPUs in its node, which amounts to FSDP's owner-rooted Broadcast confined to the node. We call it intra-node gathering, and ZeRO-3 gathers once more before the backward pass. On FSDP every parameter gathering is a cross-node broadcast from the owner, whereas here each parameter crosses nodes only once per step, at publish time. The two large AdamW parameter sets, the embedding and output layers, do not take this path and are synchronized as in plain HSDP.

\subsection{Cross-Step Publish}

The AllReduce of plain HSDP must complete both halves, reduction and return, during the backward pass. \mbox{DMuon-2}'s publish half need not stay in the backward pass. Its result is the full parameters on the leaders, which can be held until the next forward pass gathers them.

\begin{figure}[t]
\centering
\includegraphics[width=\columnwidth]{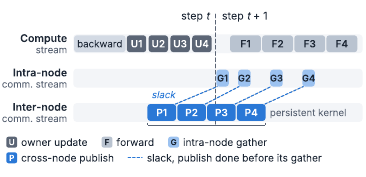}
\caption{Timeline of the cross-step publish on three CUDA streams (schematic). U is an owner update, F the next step's forward pass, P an inter-node publish, G an intra-node gather, and the dashed lines the slack between a publish finishing and its gather starting.\label{fig:3}}
\end{figure}

There are two choices for when to publish. One rides on the prefetching of the next forward pass, fetching a layer from its owner across nodes just before it is used. The other sends to the leaders right after the owner's update. Prefetching usually runs only one layer ahead, so one layer's cross-node transfer can only be hidden by the previous layer's computation, and once transferring a layer takes longer than computing one, the forward pass stalls waiting for it. \mbox{DMuon-2} therefore chooses the second option, moving the publish after the optimizer update so that it overlaps with the remaining parameter updates and with the next forward pass (Figure~\ref{fig:3}).

The publish order follows how the next step uses the parameters. Owners update parameters in the layer order of the previous forward pass and publish each parameter as soon as it is updated, without waiting for the whole optimizer step to finish. Before the next forward pass gathers a layer within the node, the leader first waits for that layer's publish to complete. A layer's publish starts when its update completes and must finish by the time the next forward pass uses it. In between lie the updates of all later layers and the forward computation of all earlier layers, so it has usually arrived long before.

\begin{figure*}[t]
\centering
\includegraphics[width=\textwidth]{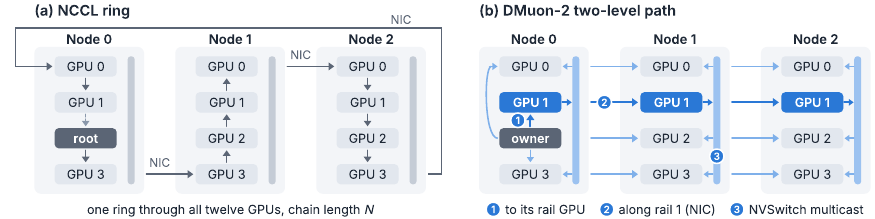}
\caption{NCCL ring versus the \mbox{DMuon-2} two-level path, on three nodes of four GPUs ($N=12$), where each row of same-numbered GPUs is a rail. (a) The ring has chain length $N$; grey marks a path that uses no NVSwitch. (b) (1) to (3) trace slice 1, and slices 0, 2 and 3 use their own rails in the same way (light blue), so each rail's chain is only $R=3$ nodes.\label{fig:4}}
\end{figure*}

\subsection{Cross-Node Traffic and Column Balancing}

The owner and the leaders of all rows lie in one column and form exactly one replicate group, and a replicate group's cross-node communication uses a single rail (Section 2.3). If both reduction and publish stayed within this group, each matrix could use only one of the $K$ rails at a time. Reduction and publish are therefore handled differently.

Reduction data comes from all ranks, and in every node all $K$ GPUs hold a gradient of $W$. \mbox{DMuon-2}'s ReduceScatterV splits the sum into $K$ slices as it computes it within the node, and each GPU accumulates one slice node by node over its own rail (Section 5). All rails thus share the reduction evenly, just as each GPU of plain HSDP synchronizes its own shard over its own rail.

Publish data has a single source, the owner, and only needs to reach the leaders of the other rows, so the new parameters are passed node by node along the owner's column rail and a single matrix occupies just one rail. Matrices of different columns publish simultaneously, and all rails carry load only if owners are distributed evenly across columns. Balancing the NS of individual ranks does not guarantee this, since some columns may receive too many large matrices and make their rails a bottleneck, so \mbox{DMuon-2}'s LPT first equalizes the NS time assigned to each column and then balances the ranks within a column.

\subsection{Communication Volume and Memory}

Across nodes, reduction and publish are the two halves split from plain HSDP's AllReduce. Each passes node by node over $R-1$ hops of $P_W$ bytes, $2(R-1)P_W$ in total, the same as plain HSDP's cross-node AllReduce of the same matrix, and intra-node parameter gathering and gradient reduction match plain HSDP's intra-node AllGather and ReduceScatter. \mbox{DMuon-2} thus adds no logical communication on HSDP either.

As for memory, plain HSDP keeps one copy of the FP32 parameters per node, which serves both as the master parameters for the optimizer update and as the source for gathering parameters in the forward pass. \mbox{DMuon-2} separates these two uses. Each node keeps only one resident BF16 copy of the parameters for intra-node gathering, each GPU storing the matrices whose owners are in its column. The FP32 master parameters, like the reduced gradients and momentum, are stored only once, on the owner. Let $M$ be the total number of elements of the Muon matrices. Per GPU on average, plain HSDP stores $4M/K$ bytes of parameters and \mbox{DMuon-2} $2M/K+4M/(RK)$ bytes. The two are equal at $R=2$, and \mbox{DMuon-2} saves more as nodes are added. The reduced gradients and momentum are only $1/R$ of plain HSDP's.

\begin{figure*}[t]
\centering
\includegraphics[width=\textwidth]{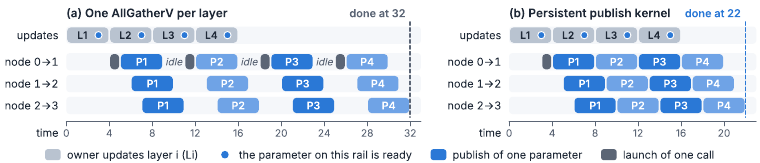}
\caption{One AllGatherV per layer versus the persistent publish kernel (schematic). One rail through four nodes, each layer publishing one parameter (P1--P4) as four tiles along the chain, both panels sharing one time scale: (a) finishes at 32, (b) at 22.\label{fig:5}}
\end{figure*}

\section{Variable-Length Communication on Two-Level Networks}

\mbox{DMuon-2} implements two variable-length collective operators, AllGatherV and ReduceScatterV, together with a persistent kernel for the cross-node publish, on NCCL's GPU-Initiated Networking (GIN). Each operator completes a layer's variable-length communication in a single kernel, reads and writes directly on the memory used by training, and occupies no more thread blocks than the corresponding NCCL primitive. The design rests on two ideas, scheduling data in the order it becomes ready (Section 5.1) and routing along the two-level network (Section 5.2). The cross-node publish builds on both as a persistent kernel (Section 5.3).

\subsection{Ready Scheduling}

Per-owner communication consists of many small and medium messages, each paying the fill and drain of a ring on its own, a cost independent of data size and set by the chain length, so it grows with the node count. NCCL has no variable-length reduction, so the Reduces in a group can only run in sequence. On the broadcast side it can merge a group's Broadcasts into one AllGatherV, but still passes data in a fixed order along a ring that strings together all ranks. Even with a layer's communication fused into one kernel, the GPU must wait for earlier data to arrive before it can process later data that is already in place. Once owners are spread over many nodes, the order of arrival no longer matches the order of processing, and this waiting grows with the number of nodes.

Each operator removes the ordering dependency between sources. Each rail keeps one queue per source node, the thread groups in the kernel handle whichever queue has data ready, local data is sent as soon as the kernel starts, and data from upstream nodes is forwarded as soon as it arrives. For a broadcast, sending can start as soon as the GPU holding the owner's data enters the kernel, without waiting for the other GPUs in its node, so all nodes holding data start sending at launch and the communication of neighbouring parameters overlaps. The more spread out the owners and the more nodes, the larger the gain. The cross-node publish follows the same principle, sending each parameter as soon as its update completes without waiting for the previous one to be delivered. Appendix C describes the queues and thread groups.

\subsection{The Two-Level Path}

Ready scheduling removes the waiting caused by a fixed order, but the fixed cost of chain length remains, and shortening the chain has to start from the path. \mbox{DMuon-2}'s AllGatherV and ReduceScatterV organize their paths around the two-level network (Figure~\ref{fig:4}). Take AllGatherV. The owner's data is sliced by rail. On the owner's node, the GPU responsible for each rail reads its slice over NVLink, writes it with one multimem store to the same location on all GPUs of the node, and at the same time sends it over its own rail to the same-numbered GPU of the next node, which likewise multicasts it within its node and passes it on. ReduceScatterV runs in the opposite direction. On every node the GPU responsible for a rail reads the sum of the same gradient segment over all GPUs of the node with one multimem reduction load, adds the partial sum from upstream, and sends it to the next node, the last hop writing directly into the owner's memory. Intra-node multicast and reduction each take a single hop, so the chain length drops from the number of ranks $N$ to the number of nodes $R$, 1/8 of an NCCL ring on 32 nodes. The cost is somewhat more traffic inside the node, since intra-node multicast and reduction route even the source GPU's own share out through the NVSwitch and back, taking about 1/8 more intra-node bandwidth, while the logical volume is unchanged. As training scales out, however, the bottleneck is the hop-by-hop chain across nodes, and this extra traffic shows only for very large messages on few nodes. The per-layer volumes and node counts of real training fall in the range where the two-level path beats NCCL (Section 7.4).

\begin{figure*}[t]
\centering
\includegraphics[width=\textwidth]{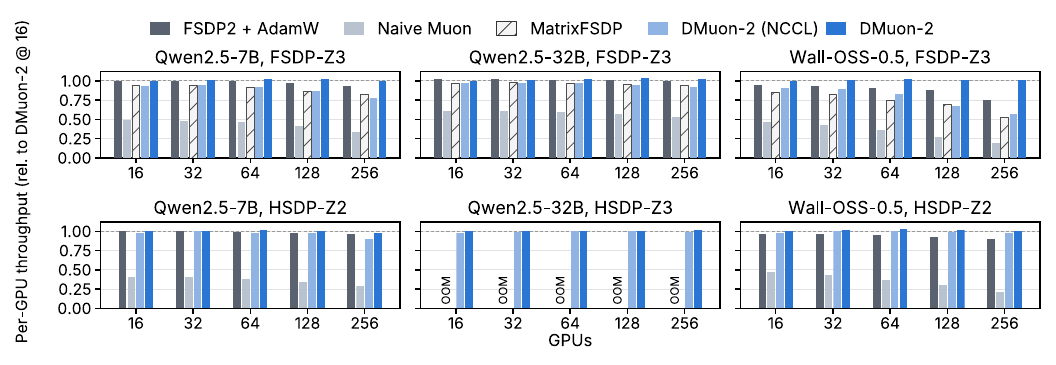}
\caption{End-to-end per-GPU throughput, normalized in each panel to \mbox{DMuon-2} on 16 GPUs. The per-GPU batch is fixed, so throughput is inversely proportional to step time. The bars at each GPU count are A, B, C, and D; the FSDP-Z3 row adds a bar M, which is MatrixFSDP~\citep{gao2026matrixfsdp}. The top row is FSDP-Z3 and the bottom row HSDP, which is HSDP-Z3 for Qwen2.5-32B, where A and B run out of memory. Appendix Table~\ref{tab:2} gives all configurations of A to D, including the two models not shown here, and Table~\ref{tab:8} those of M.\label{fig:6}}
\end{figure*}

\subsection{A Persistent Kernel for the Cross-Step Publish}

The cross-node publish has a different load from the other two operators. When parameter gathering or gradient reduction starts, the whole layer is already in place, whereas publish tasks appear one by one during the optimizer step, one each time a parameter finishes updating, and all of them fall on the single rail of the owner's column, with no intra-node multicast to use (Section 4.3). Both existing ways of sending leave this rail idle (Figure~\ref{fig:5}a). A per-layer coalesced call usually finds only one parameter of the layer on the rail, degenerates into a single-root broadcast, and has to wait for the whole layer to be updated, so the chain is refilled and drained for every layer. Per-parameter calls leave the rail idle between the end of one and the start of the next.

\mbox{DMuon-2} therefore uses a persistent kernel that chains the parameters landing on the same rail within a step into one continuous stream (Figure~\ref{fig:5}b). The kernel starts when the first publish task arrives and exits once all parameters are handled. Each time an owner finishes updating a parameter, the kernel writes its BF16 copy to the leader of the next node, every leader forwards as data arrives, and several parameters are in flight at once, so the chain fills once per step rather than once per parameter. The NIC writes the data directly between the copies on successive nodes, and the threads only issue writes and wait on signals, so a single thread block saturates a rail, whereas NCCL moves the data with SMs and uses several thread blocks for the same operation.

\section{Implementation}

\mbox{DMuon-2}'s runtime is about 16 thousand lines of Python covering owner assignment, the FSDP2 integration, and the optimizer. The device-side code of the symmetric pool and the three communication kernels is about 8.4 thousand lines of CUDA and C++. NS iterates on the Gram matrix~\citep{zhang2026gram}, and its symmetric products choose between dedicated kernels and cuBLAS by matrix shape (Appendix C.3).

\mbox{DMuon-2}'s hooks sit at the boundaries of FSDP units, that is, at every layer. Before a layer is used in the forward pass, one AllGatherV gathers its parameters into a buffer on every rank, from which the forward pass reads directly. Once a layer's backward pass finishes, one ReduceScatterV reads the gradients on all ranks and reduces each parameter's gradient to its owner. While the current layer computes, the next layer's communication is already running on another CUDA stream. The optimizer creates state and performs updates only on owners. An owner holds the whole matrix together with all of its optimizer state, the update reads only local data, and substituting another matrix optimizer does not touch the communication path.

The communication kernels read and write directly on the memory used by training, which requires the participating data to stay resident in registered windows and to sit at the same offset on every GPU, so \mbox{DMuon-2} maintains a symmetric pool with an identical layout on every GPU (Appendix C.1).

\section{Evaluation}

\subsection{Setup}

Experiments run on up to 32 nodes, each with eight H100 80GB GPUs and one 400 Gb/s RDMA NIC per GPU. The evaluation covers three dense LLMs, Qwen2.5-1.5B, 7B, and 32B~\citep{yang2024qwen25}, and two VLA models, Wall-OSS-0.5~\citep{yu2026walloss} and $\pi_{0.5}$~\citep{pi2025pi05}. FSDP shards across all GPUs, and HSDP shards within a node and replicates across nodes. Combined with ZeRO-2 or ZeRO-3, they give the three configurations used for most models: FSDP-Z2, FSDP-Z3, and HSDP-Z2. Qwen2.5-32B fits on a GPU only with its parameters sharded, so it uses FSDP-Z3 and HSDP-Z3.

We compare four lines. D is the full \mbox{DMuon-2}, A, B and C are its baselines. A is PyTorch FSDP2 with AdamW, which tests whether \mbox{DMuon-2} offsets the extra cost of Muon over an element-wise optimizer. B implements Muon along the first route~\citep{liu2025muon}, gathering whole matrices and running NS on every GPU, and represents a naive Muon implementation. C uses the same matrix owners as D but keeps NCCL for communication, isolating the gain of the operators in Section 5. The four lines A, B, C and D use the same sharding and per-GPU token count in a given configuration. M is the concurrent MatrixFSDP~\citep{gao2026matrixfsdp}, whose design point matches C. Its published experiments cover ZeRO-3 only, so we compare it under FSDP-Z3 alone, on the same sharding, model and per-GPU batch (Appendix D.9). Appendices D.1 to D.3 give the training configurations, implementation details, and measurement methods.

\subsection{End-to-End Throughput}

\mbox{DMuon-2} roughly matches AdamW and pulls ahead at scale (Figure~\ref{fig:6}). Across configurations, the step time of C is 4.6\% longer than that of A at the median, which is the overhead of Muon over an element-wise optimizer, and D offsets it. Excluding the five configurations where A's memory is close to the limit, D is no slower than A in 50 of the remaining 60 and at most 2.2\% slower in the rest. On 256 GPUs, the throughput of D is a median of 1.05 times and at most 1.34 times that of A.

\mbox{DMuon-2} is faster than naive Muon in every configuration (Figure~\ref{fig:6}). D trains 1.33 to 5.28 times as fast as B, with a median of 2.09, and in no row does the ratio shrink as the GPU count grows. B's NS work per GPU is fixed, while D divides it among more GPUs. D's extra memory over C is the symmetric pool, 1.0 to 7.5 GiB (Appendix D.4).

The gain of the communication operators grows with scale (Figure~\ref{fig:6}). D's median gain over C rises from 3.2\% on 16 GPUs to 13.9\% on 256 and reaches 78\% at most. The gain depends on the sharding, with medians of 11.2\% under FSDP-Z3, 6.7\% under FSDP-Z2, and about 1.6\% under HSDP. It also depends on the model, with medians of 9.8\% on Wall-OSS-0.5 and 3.2\% on Qwen2.5-32B. Wall-OSS-0.5 communicates many small matrices and spends little time computing each layer, so its sparse variable-length collectives are largely exposed on the critical path, which is where the operators help most. HSDP gathers parameters layer by layer only inside a node, where NCCL is already efficient, so the two variable-length operators gain little. On HSDP the cross-node cost is carried by the persistent publish kernel, which occupies one thread block per rail, interferes little with the compute kernels it overlaps, and moves the publish off the critical path of the next forward pass (Section 7.3).

FSDP-Z3 also allows a comparison with the concurrent work. Across the 20 FSDP-Z3 configurations of four models on which M was run (Table~\ref{tab:8}), D is faster than M in every one, by 1.04 to 1.93 times with a median of 1.09. The median margin of D over M grows from 1.05 on 16 GPUs to 1.21 on 256 and is largest on Wall-OSS-0.5 (Figure~\ref{fig:6}, top row). Appendix D.9 gives M's configuration and tuning.

\subsection{Globally Unique State and Cross-Step Publish}

\begin{figure}[t]
\centering
\includegraphics[width=\columnwidth]{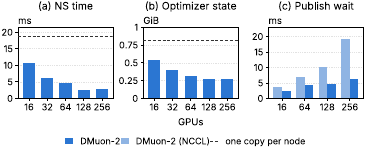}
\caption{Globally unique owners and the cross-step publish, by GPU count. (a) NS kernel time of the slowest rank and (b) optimizer state on the busiest rank, for Qwen2.5-1.5B under HSDP-Z2. Bars are \mbox{DMuon-2}, with owners assigned across the cluster and a single copy of the state. The dashed lines are the placement with one copy per node, estimated in (a) and computed in (b). (c) Wait before the forward pass for publishing, Wall-OSS-0.5 under HSDP-Z2. Values are in Appendix Tables 3 and 4.\label{fig:7}}
\end{figure}

Globally unique owners make NS time and optimizer-state memory shrink with the GPU count (Figure~\ref{fig:7}a, b). For Qwen2.5-1.5B under HSDP-Z2, the NS kernel time of the slowest rank falls from 10.7 ms on 16 GPUs to 4.5 ms on 64 GPUs, following the number of matrices per rank, which drops from 13 to 5. With owners assigned only within a node, as in plain HSDP, the estimate from the same per-shape NS cost is about 18.8 ms; this is a lower bound and does not change with the node count. The optimizer state behaves the same way, at 0.27 GiB on the busiest rank on 128 GPUs against a constant 0.81 GiB with one copy per node (Appendix Table~\ref{tab:3}).

The cross-step publish moves parameter publishing off the critical path of the next forward pass (Figure~\ref{fig:7}c). For Wall-OSS-0.5 under HSDP-Z2, the wait before the forward pass for publishing falls from 10.2 ms under C to 4.8 ms on 128 GPUs and from 19.3 ms to 6.3 ms on 256 GPUs (breakdown in Appendix Figure~\ref{fig:10} and Table~\ref{tab:4}).

\subsection{Variable-Length Communication Operators}

\begin{figure*}[t]
\centering
\includegraphics[width=\textwidth]{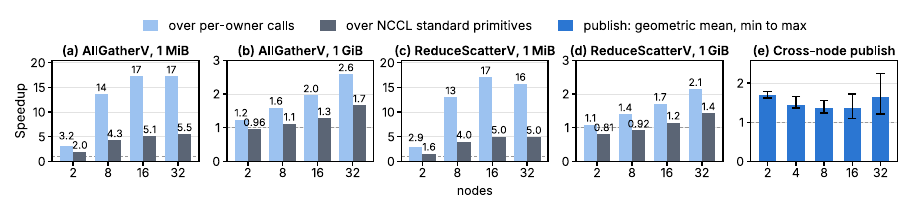}
\caption{Speedup of the three communication operators over their baselines, with the node count on the horizontal axis. Values above 1 (dashed line) mean ours is faster. (a) to (d) give the two variable-length operators on 1 MiB and 1 GiB messages, light blue against per-owner calls and dark gray against the faster standard NCCL primitive at the same message size. (e) gives the persistent publish kernel against the faster of per-owner and coalesced calls on one real rail of each of the five models, with bars at the geometric mean and whiskers from the minimum to the maximum. Values are in Appendix Tables 5 to 7.\label{fig:8}}
\end{figure*}

Both variable-length operators are faster than the general implementation for sparse variable-length layouts and also than NCCL's standard primitives, which handle only regular layouts, and the margin grows with the node count (Figure~\ref{fig:8}a to d). For messages of 1 MiB and below, the gain comes from fusing a layer's variable-length communication into one kernel, which issues one call where per-owner calls issue one per owner. On 8 to 32 nodes these messages are 6.9 to 17.3 times as fast as per-owner calls and 3.3 to 6.3 times as fast as the standard primitives. For 1 GiB messages, the gain comes from the two-level path, which shortens the chain from the number of ranks to the number of nodes, but intra-node multicast and reduction take about 1/8 more intra-node bandwidth (Section 5.2). These messages are therefore slightly slower than the standard primitives on 2 nodes (0.81 to 0.96 times), overtake them from 16 nodes, and are 1.43 to 1.66 times as fast on 32 nodes. Real training mostly moves 256 MiB to 1 GiB per layer, a range in which our operators are faster than the standard primitives from 16 nodes on.

The persistent publish kernel is also faster on real loads (Figure~\ref{fig:8}e). Replaying all publish tasks of a step on one rail of each of the five models, it is 1.09 to 2.25 times as fast as the faster of per-owner and coalesced calls on 2 to 32 nodes. On 32 nodes it exceeds 2 times for Qwen2.5-1.5B and Wall-OSS-0.5, because per-owner calls refill and drain the chain for every parameter. The largest load, Qwen2.5-32B, is bandwidth-bound and gains the least, 1.09 to 1.23 times from 8 nodes on.

\subsection{Numerical Equivalence}

Every timed run of the communication operators checks its result: the gathered parameters equal the owner's data bit for bit, and the reduction of integer inputs equals the exact sum bit for bit. Reduction rounds to BF16 at every level, and on 64 and 128 GPUs its maximum relative error against the float64 exact sum is 0.0096 to 0.0165, and at every message size it is less than half of NCCL's ring reduction.

D and the reference implementation B remain equivalent over a long run: each was trained for 10000 steps on two models, B twice with different seeds (Appendix Figure~\ref{fig:11}). On Qwen2.5-1.5B with FineWeb-Edu text~\citep{penedo2024fineweb} the step-0 loss is bit-identical between the two, the median per-step relative difference over the remaining steps is 0.0028\%, and the final losses are 6.175 and 6.168; rerunning B with a different random seed gives 0.1022\% for the same median. Training of Wall-OSS-0.5 is not bitwise deterministic; the median per-step relative difference is 3.12\% when the implementation changes and 2.98\% when only the seed changes. Neither difference grows with step. The configuration and a short-horizon check over 13 further configurations are in Appendix D.8.

\section{Related Work}

\textbf{Matrix-granular sharding on FSDP.} veScale-FSDP~\citep{wang2026vescale} places whole blocks on single devices to support block-wise quantization, matrix optimizers, and zero-copy communication; like the FSDP version of Canzona released later~\citep{wang2026canzona}, it picks one rank per tensor, gathers the matrix there, runs NS, and sends the result back. The concurrent MatrixFSDP~\citep{gao2026matrixfsdp} places each matrix whole on one owner in a ZeRO-3 layout, but the replicate dimension is 1 in all of its published experiments. \mbox{DMuon-2} leaves FSDP2 unchanged, reduces gradients to owners during the backward pass, and on HSDP handles owners across replicas, a single cluster-wide copy of the optimizer state, and the parameter publish.

\textbf{Hybrid sharding for element-wise optimizers.} The hybrid sharding of Megatron-FSDP, HFSDP~\citep{nvidia2026megatronfsdp}, also shards the optimizer state across nodes and gathers the updated parameters during the first forward pass of the next step. In \mbox{DMuon-2} owners hold whole matrices, so the optimizer step needs no gather, and parameters are published as soon as they are updated, so they arrive earlier (Section 4.2).

\textbf{Owners for matrix optimizers.} ZeRO~\citep{rajbhandari2020zero} partitions the optimizer state, and Distributed Shampoo~\citep{shi2023shampoo} assigns the preconditioners of different blocks to different ranks. For Muon, the layer-wise distributed optimizer of Megatron-LM~\citep{nvidia2026megatronmuon} and Canzona~\citep{wang2026canzona} assign whole parameters to ranks, while Kimi K3~\citep{kimi2026k3} and GLM-5~\citep{glm2026glm5} have each rank fetch only the parameters or shards it is responsible for. Together with Step 3.5 Flash and DeepSeek-V4 in the introduction, these systems follow the second route, in which parameters are fully replicated during the forward and backward passes. \mbox{DMuon-2} targets FSDP and HSDP, where parameters are also sharded. Our team's technical report describes owner-based Muon~\citep{chen2026dmuonreport}.

\textbf{Topology-aware collective communication.} BlueConnect~\citep{cho2019blueconnect}, Blink~\citep{wang2020blink}, SCCL~\citep{cai2021sccl}, TACCL~\citep{shah2023taccl}, TCCL~\citep{kim2024tccl}, SyCCL~\citep{cao2025syccl}, and HeteCCL~\citep{hei2026heteccl} decompose or synthesize collective schedules for a given network hierarchy; MSCCLang~\citep{cowan2023msccllang} provides a language for them; and HiCCL~\citep{hidayetoglu2024hiccl} composes collectives level by level from multicast, reduction, and synchronization primitives. These systems mainly target collectives in which every rank contributes the same length and all data is in place at the start. In \mbox{DMuon-2} segments differ in length and root, and the publish data is produced parameter by parameter by the optimizer, so it has to be scheduled in the order it becomes ready (Section 5.1).

\section{Conclusion}

Data-parallel runtimes shard model state by element, while optimizers such as Muon update whole matrices, and this mismatch forces existing implementations to trade extra communication for repeated computation. \mbox{DMuon-2} lets the unit of distribution follow the unit of the optimizer. Every matrix has a single owner in the cluster, its NS runs once and its state is stored once, and the resulting sparse, variable-length communication is organized around the two-level network and the order in which data becomes ready. Data parallelism has always decided, on the optimizer's behalf, how its state is partitioned. As optimizers begin to update parameters a whole matrix at a time, we believe that decision should rest with the structure of the optimizer itself.

\bibliographystyle{mlsys2025}
\bibliography{refs}

\section*{AI Usage Disclosure}

We used Anthropic Claude, a large-language-model assistant accessed through the Claude Code command-line tool, for writing, figures, and parts of the engineering work.

The research questions, system design, and arguments are the authors' own. This includes the matrix-granular ownership framework, the globally unique state and cross-step publish on HSDP, the ready scheduling and two-level path of the variable-length operators, and the choice of experiments and the conclusions drawn from them.

For writing, the authors set the claims, structure, and data of each section, and the assistant drafted, revised, and polished the text to their instructions. The authors reviewed all text paragraph by paragraph. For figures, the assistant wrote plotting scripts to the authors' specification, and every figure reads its data directly from the experimental results.

For code, the authors designed the architecture, interfaces, and key mechanisms of the \mbox{DMuon-2} runtime and its communication operators, and the assistant implemented the code under this design, which the authors reviewed and tested. For experiments, the assistant drove experiment scripts, submitted and monitored cluster jobs, and parsed timing output. All data in this paper come from real runs on the hardware described, and no value was generated, completed, or inferred by the model.

For related work, the assistant searched for and checked references, and every entry in the bibliography was verified against its source.

The assistant is not an author. The authors take full responsibility for the correctness, originality, and integrity of the paper.

\appendix
\makeatletter
\setlength{\@fptop}{0pt}\setlength{\@fpsep}{16pt plus 4pt}\setlength{\@fpbot}{0pt plus 1fil}
\setlength{\@dblfptop}{0pt}\setlength{\@dblfpsep}{16pt plus 4pt}\setlength{\@dblfpbot}{0pt plus 1fil}
\makeatother
\setcounter{dbltopnumber}{4}\renewcommand{\dbltopfraction}{.95}\renewcommand{\dblfloatpagefraction}{.4}
\section{The Criterion for Sharded Orthogonalization}

This appendix derives the result stated in Section 2.1. The criterion below concerns the orthogonal factor obtained from an exact polar decomposition.

Dropping the subscript $t$ and assuming $M$ has full row rank, $UV^\top=(MM^\top)^{-1/2}M$. FSDP cuts $M$ into row blocks along dimension 0. With two shards $M_1$ and $M_2$, orthogonalizing each shard separately gives $Q_i=(M_iM_i^\top)^{-1/2}M_i$, and
\begin{equation*}
Q_1Q_2^\top=(M_1M_1^\top)^{-1/2}\,M_1M_2^\top\,(M_2M_2^\top)^{-1/2}.
\end{equation*}

$UV^\top$ is semi-orthogonal, so any two of its rows have zero inner product, and the product of its upper and lower blocks must be the zero matrix. For the concatenation to equal $UV^\top$ we therefore need $Q_1Q_2^\top=0$, i.e., $M_1M_2^\top=0$; conversely, when $M_1M_2^\top=0$, $MM^\top$ is block diagonal and the concatenation equals $UV^\top$ exactly. Hence sharded orthogonalization agrees with whole-matrix orthogonalization if and only if rows of different shards are pairwise orthogonal.

The case of more shards is analogous. When $m>n$, $UV^\top=M(M^\top M)^{-1/2}$, where $M^\top M=\sum_i M_i^\top M_i$ aggregates all shards, and the conclusion is unchanged.

\section{Cost Model and Greedy Algorithm for Owner Assignment}

This appendix gives the details omitted from Section 3.3.

\textbf{The assignment problem.} Let $o(u)$ be the owner shared by the matrices of assignment unit $u$, let the NS time $t_u$ and the byte count $P_u$ of $u$ be the sums over its Muon matrices, and let $\mathcal{U}_\ell$ be the set of assignment units of layer $\ell$. Owner assignment can be written as the optimization problem
\begin{align*}
&\min_{o}\ \max_{j}\sum_{o(u)=j} t_u \\
\text{s.t.}\ &\sum_{o(u)=j} P_u \le (1+\epsilon)\,\frac{P_M}{N}\ \ \forall j \\
\phantom{\text{s.t.}}\ &o(u)\ne o(v)\ \ \forall \ell,\ \forall u\ne v\in\mathcal{U}_\ell
\end{align*}
that is, minimize the NS time of the busiest rank subject to every rank $j$ holding at most $1+\epsilon$ times the average bytes and the assignment units of the same layer belonging to different ranks. The four models in Figure~\ref{fig:2} have only 1 to 6 assignment units per layer, fewer than the number of ranks, so the layer-spreading constraint can always be met. The byte cap cannot always, and with large assignment units the problem may even be infeasible.

\textbf{Why NS time is measured.} In Qwen2.5-1.5B, the k and v projections have only 1/35 of the elements of the MLP projections, yet take 40--50\% of their NS time. Prior work estimates cost by element count in its experiments~\citep{wang2026canzona}. Replacing $t_u$ with element count, the busiest of 32 ranks for Qwen2.5-1.5B spends 1.81 times the average NS time, no better than 1.74 without balancing. Estimating by $m^2n$ is also inaccurate, since smaller matrices are dominated by fixed overheads, and because \mbox{DMuon-2} iterates on the Gram matrix by default, the relative time of elongated matrices is far below this estimate.

\textbf{Why the byte cap is kept.} Assigning by time alone, the heaviest of 32 ranks for Qwen2.5-1.5B holds 1.34 times the average bytes. With $\epsilon=0.1$, as in \mbox{DMuon-2}, this drops to 1.14, while the NS time of the busiest rank only moves from 1.10 to 1.14 times the average.

\textbf{The greedy in detail.} All assignment units of the model are sorted by $t_u$ in decreasing order, with ties broken by element count and then parameter name, and each in turn is given to one rank. To choose the rank, \mbox{DMuon-2} first excludes ranks already used in the unit's layer, and among the rest picks the one with the smallest accumulated time whose byte count would stay under the cap after receiving the unit. If every candidate would exceed the cap, it picks the one with the fewest bytes.

\textbf{Measurement protocol.} Every rank measures the NS time of each shape on its own GPU, and after aggregation each shape takes the median across ranks, so measurement noise on a single GPU hardly changes the assignment. Every rank computes from the same aggregated data and obtains the same assignment. The NS times in Figure~\ref{fig:2} are measured on H100 with \mbox{DMuon-2}'s default Gram iteration, and the times used for assignment and for evaluation come from two independent measurements, each taking the median of 8 GPUs, with every point in the figure obtained by accumulating these times according to the assignment rather than by timing a whole run.

\textbf{Effect of the layer-spreading constraint.} Without layer spreading, on 16 ranks, in the worst layer of each of the four models the heaviest root carries 1.06 to 2.4 times the data of the layer's largest assignment unit, and with it this ratio drops to 1, at the cost of the busiest rank's NS time rising at most from 1.01 to 1.11 times the average (Qwen2.5-7B, 32 ranks).

\textbf{Effect of column balancing.} Without column balancing, on 16 nodes the heaviest rail of $\pi_{0.5}$ carries 3.17 times the average publish bytes, and that of Qwen2.5-7B 1.33 times. With it, all four models stay within 1.06 times.

\section{Implementation Details}

This appendix gives the implementation details omitted from Section 6.

\subsection{The Symmetric Pool}

The AllGatherV and ReduceScatterV kernels read and write the memory holding parameters and gradients directly through NCCL's device-side API~\citep{hamidouche2025gin}. This requires the data to reside in registered windows, windows used for multicast must be bindable to a multicast object, and the same data must sit at the same offset on every GPU. Memory allocated through PyTorch's default \texttt{cudaMalloc} path meets none of these conditions. The caching allocator places parameters and gradients at different addresses every step, which also differ across GPUs, and registering memory as a multicast-capable window is a collective operation whose cost is independent of size, about 20 ms on our platform, so it cannot be done on the fly for every communication. Copying data into a registered buffer before communicating costs every GPU an extra copy of its own data each time, consuming bandwidth for large messages and adding latency for small ones. \mbox{DMuon-2} therefore fixes the location of the parameters and gradients that take part in communication, keeping them resident in registered memory. Keeping communication data resident in fixed buffers to avoid copy-in and copy-out is not new. veScale-FSDP maps each tensor to a fixed position in a global communication buffer (Section 8), and Megatron-FSDP also registers its parameter and gradient buffers as NCCL symmetric windows so that NCCL can use more SM-efficient algorithms~\citep{nvidia2026megatronfsdp}. \mbox{DMuon-2}'s pool serves its own kernels, which multicast and write across nodes directly in the variable-length owner pattern. The offsets of parameters and gradients are planned in execution order and reused across groups, and gradients produced by autograd must land in the pool without changing model code.

\begin{figure*}[t]
\centering
\includegraphics[width=\textwidth]{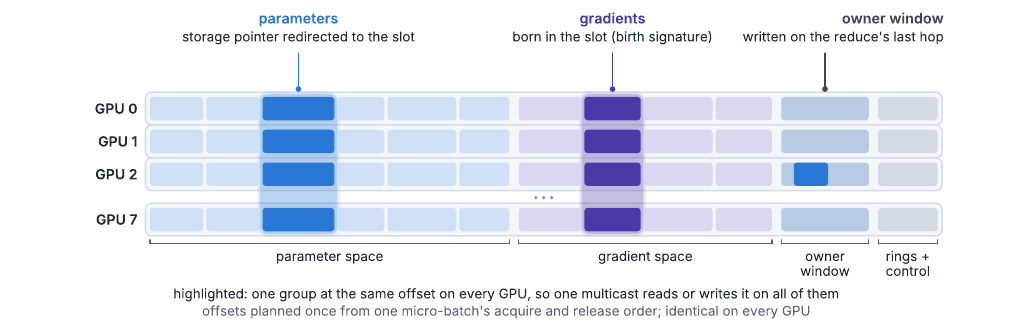}
\caption{The symmetric pool. The pool has the same layout on every GPU, consisting of the parameter space, the gradient space, the owner window, and the rings and control area used by the kernels. The highlighted parameter group and gradient group sit at the same offset on every GPU, so one multicast reads or writes the group on all GPUs. Parameters enter their slots by redirecting the storage pointer, gradients are born in their slots through their allocation signatures, and the last hop of the reduction writes the owner window directly. The dark cell on GPU 2 is its part as an owner. Offsets are planned once from the acquire and release order of one micro-batch and are identical on all GPUs.\label{fig:9}}
\end{figure*}

On every GPU \mbox{DMuon-2} allocates a parameter space, a gradient space, and an owner window with identical layouts and registers them as NCCL memory windows, together called the symmetric pool. The parameters and gradients handed to \mbox{DMuon-2} are taken from the pool at fixed offsets (Figure~\ref{fig:9}). Offsets are laid out by the order in which the groups' parameters and gradients are acquired and released in a steady micro-batch, and freed ranges are reused by later groups. After training starts, the first micro-batch only warms up, the second communicates as usual and records the acquisition order, and the pool opens before the first optimizer step. All later communication runs on the pool. All GPUs execute the same sequence of hooks, so the offset tables they compute independently agree by construction, and a digest exchange at pool opening checks this once.

Parameters are already views into packed buffers, so gathering only needs to point the buffer's storage at its position in the pool. The parameters, the views owners write through, and the weights the forward pass saves for the backward pass all follow to the new address, and model code is unchanged. Gradients are allocated by autograd itself. \mbox{DMuon-2} wraps PyTorch's caching allocator and records a allocation signature for every allocation in the backward pass, consisting of the layer, the index of the autograd node within the layer, the index of the allocation within the node, and the byte count. During recording it finds which allocation is finally handed to each parameter, and from then on a matching signature returns the corresponding address in the pool directly, so the gradient is in the pool the moment it is produced. Identifying gradients by operator type cannot achieve this. When the same linear layer is used twice, autograd allocates a separate buffer to sum the two gradients, and the final gradient of a normalization layer comes from a type-cast node. Only tracing the allocation back from the memory finally delivered to the parameter is independent of operators. In real training the signature hit rate exceeds 99.99\%. On the rare mismatch, the next micro-batch relearns, and in the current step the gradient is first copied into the pool and then reduced. Each owner's window holds the reduced gradients of the matrices it owns. The last hop of the reduction writes into it directly, and the optimizer reads from it without another copy.

The cost is that the gradient space and the owner window become resident instead of being freed after use. Training Qwen2.5-7B with ZeRO-2 on 2 nodes, the two together take about 1.7 GiB, and the owner window shrinks as GPUs are added. With the fixed overhead of communication contexts included, the peak memory of Qwen2.5-7B and Wall-OSS-0.5 on 4 nodes is 1.7 and 1.2 GiB higher than before, about 2\% of a GPU's memory. Under ZeRO-2 the parameters are kept from the forward to the backward pass anyway, so placing them in the pool costs no extra memory. The few groups that cannot be placed in the pool, such as groups containing a parameter that never receives a gradient, still use grouped communication. On HSDP the intra-node gathering uses the same AllGatherV kernel, with the node-local shard group as its communicator. The whole path can be turned off as a unit. Each matrix then still runs NS only once, and communication falls back to grouped calls.

\subsection{Communication Kernels}

\textbf{Queues and thread groups of ready scheduling.} Data is cut into 64 KiB chunks and each rail has an independent queue per source node. A queue is just a ring buffer plus a few counters. When there are no more source nodes than thread groups, each group serves one queue. With more source nodes, each group serves its queues in turn, handling whichever has data ready. The waiting, multicast, and sending of each chunk are mostly latency, so several thread groups must work on different chunks at once to saturate a rail's bandwidth. The number of thread groups is bounded by the SM budget, while the number of queues grows with the number of nodes, so different sources never have to share a queue. Data within a queue arrives in chain order, and both ends use the ring-buffer slots in the same sequence, so chunk positions need no per-chunk negotiation. A thread group serving several queues never blocks on a queue without ready data, and dependencies only flow one way, from owner nodes downstream, so no deadlock can form.

\textbf{Publish copies and version checking.} The leaders' BF16 copies live in registered memory with fixed positions for the parameters of each rail, and the publish passes through no staging buffer. Every parameter's arrival flag carries the step number, and before the next forward pass gathers a layer, the leader checks the flag and never mistakes the previous step's version for the new one.

Intra-node multicast and reduction are issued with the multimem family of PTX instructions. One \texttt{multimem.st} writes data to all GPUs and one \texttt{multimem.ld\_reduce} reads back the sum of the same location on all GPUs~\citep{nvidia2026cuda}.

AllGatherV, ReduceScatterV, and the cross-node publish of HSDP (Section 5.3) are each a single kernel written with the device-side API. For the first two, the kernel issues intra-node multicast and reduction directly against the same offset in every GPU's pool. For all three, the kernel itself initiates the RDMA that writes cross-node chunks to the next node and updates the peer's signal, with no CPU proxy thread moving data and no relay of multiple kernels. A group's segment table is frozen when the group is built. Each call passes only the base address leased for that call, and the kernel slices and assigns queues by the table without regenerating tasks. The thread-block count has a fixed cap. Each block has 256 threads. Large gathering and reduction groups use at most 16 blocks, the actual count following the number of pipelines that carry data, no more than NCCL's equal-length primitives use for the same total bytes. With at most 16 nodes and little data per node, the source node instead reaches every node in one hop, which we call the star path, and the short segments that take it occupy a single SM. Exceeding the cap is an error at group construction.

A cross-node write is complete when it is visible to the peer. Data and signals are written in release/acquire order, so when the receiver sees the signal the data has certainly arrived. Each queue of the two operators has a ring buffer of only a few slots at the receiver, and the sender must wait until the receiver acknowledges the chunk that last occupied a slot before writing the next one into it. Arrival counters, acknowledgement counters, and send/receive sequence numbers live in registered memory and increase monotonically across calls, never reset. All GPUs issue calls in the same order, so the sequence numbers at both ends advance in lockstep, and a call need not wait for the last acknowledgement or drain the ring buffer when it ends.

HSDP uses ReduceScatterV to aggregate gradients across nodes from all data-parallel ranks onto global owners. Parameter gathering instead uses one AllGatherV within each node. Its communicator spans a single node and uses only NVSwitch, so the gain comes from fusing a layer into one kernel and from multicast rather than from a shorter cross-node chain. In addition, every leader also holds a BF16 copy of the parameters, from which its node's gathering is issued. Each time an owner finishes updating a parameter, it rounds the FP32 master parameters to BF16, writes them into its own copy, and hands them to the cross-node publish. The publish kernel requires these copies to reside in registered memory with a fixed layout on each rail, and \mbox{DMuon-2} brings it in in two steps. After the optimizer is built and before the first forward pass, \mbox{DMuon-2} moves each leader's copy into a block of registered memory, fixing the layout on each rail, and training then runs directly on this memory. The warm-up steps still publish with grouped communication. Once warm-up ends, \mbox{DMuon-2} drains the in-flight broadcasts, builds the publish kernel's communicator and windows, and the kernel takes over all later publishes. The training script takes no part in this process, and the FP32 master parameters always stay in memory managed by PyTorch. This memory is allocated symmetrically by the heaviest rail. Training Qwen2.5-7B with HSDP on 4 nodes, each GPU uses about 1.25 GiB more.

\subsection{Gram Newton--Schulz and Its Fast Kernels}

\mbox{DMuon-2}'s NS uses Gram Newton--Schulz~\citep{zhang2026gram}. Every NS step left-multiplies the matrix by a polynomial of the Gram matrix $XX^\top$, so Gram NS moves the iteration onto the smaller Gram matrix and multiplies back into the original matrix once at the end, which is equivalent to the original iteration in exact arithmetic. The square matrices in the iteration commute with each other and their products remain symmetric, so only the lower triangle needs to be computed and then mirrored. The per-step coefficients, the mid-way restart, and the mixed-precision flow follow the reference implementation, with the five-step Polar Express coefficients~\citep{amsel2025polar}.

On Hopper, symmetric products use the kernels of QuACK, an open-source kernel library written in CuTe DSL~\citep{guo2025quack}. QuACK's symmetric-product kernels support only Hopper and newer architectures, and \mbox{DMuon-2} provides its own symmetric-product kernel for Ampere, also written in CuTe DSL. Both are optional dependencies, and without them \mbox{DMuon-2} falls back to cuBLAS's general matrix multiplication.

Symmetric kernels are not faster than cuBLAS for every shape, and one NS consists of twenty to thirty small kernels, so \mbox{DMuon-2} does two things per shape at runtime. The first is kernel selection. When a shape's symmetric product first appears, \mbox{DMuon-2} checks its result against cuBLAS, discards configurations whose error exceeds the limit, and then times the remaining configurations against cuBLAS. The winner is cached on disk per GPU model, and later training dispatches by table lookup. The object files compiled from the CuTe DSL kernels are also cached on disk, so restarting training needs no recompilation. The second is graph capture. A rank runs NS on dozens of matrices per step, adding up to more than a thousand kernel launches, yet usually only a dozen or so shapes. \mbox{DMuon-2} captures the whole NS into a CUDA Graph the second time each shape appears and only replays it afterwards. The first time runs normally and also completes the timing and compilation above. Replay has almost no host-side cost, which is why kernel selection times only execution on the GPU. Learning rate and weight decay stay outside the graph, so changing them needs no recapture.

\subsection{Combining with Activation Recomputation and Tensor Parallelism}

Activation recomputation re-runs a layer's forward pass inside the backward pass, and \mbox{DMuon-2}'s per-layer hooks detect this by checking whether an autograd backward task is active. A forward that runs inside backward does not prefetch, does not record layer order, and does not reshard afterwards; all three hold only in a true forward pass, since the layer to prefetch is the one backward will consume next, the order was recorded during the true forward, and the parameters are about to be used by this layer's backward. Every Qwen2.5-32B run in this paper enables full recomputation.

Tensor parallelism is supported through DTensor, covering its combination with both FSDP2/ZeRO-2 and HSDP. \mbox{DMuon-2} takes no mesh argument and instead infers whether a parameter is sharded from its DTensor mesh and placements. A sharded matrix therefore has two owners: the data-parallel owner of Section 3, and one along the tensor-parallel axis chosen by the same greedy algorithm within each data-parallel owner's bucket. Within a step, the local gradients of the tensor-parallel ranks are gathered to that owner along this axis, Newton--Schulz runs on the complete logical matrix, and the update is scattered back along the same axis. Orthogonalizing the complete matrix preserves Muon's definition (Section 2.1 and Appendix A), which is where the gather and the scatter come from. The scatter completes before the cross-node publish and shares its stream, so ranks along the replicate axis only ever read correct shards. The experiments in this paper focus on scaling along the data-parallel axis and do not use tensor parallelism.

\subsection{Combining with MoE Expert Weights}

\textbf{Expert weights at two granularities.} Both the forward and the backward pass of the experts rely on a grouped GEMM, which requires the weights of a layer's experts to be merged into one three-dimensional tensor, while the unit a matrix optimizer acts on is a two-dimensional matrix. \mbox{DMuon-2} introduces a fused object for expert weights that puts both granularities on the same storage: every expert is registered as its own two-dimensional parameter, and owners, NS, and the parameter publish all work at that granularity, while the three-dimensional tensor the computation sees is a view of the same buffer. That buffer is also the destination of the parameter publish: every owner writes its updated parameter at the offset it occupies in the fused layout, so the change of form from separate experts to a complete weight is done by the publish itself. The gradient direction is the reverse: a layer's gradient arrives as one fused buffer, is first split by offset into an independent gradient tensor per expert, and is then reduced to each expert's own owner. NS still works on one 1408x2048 matrix per expert, and the orthogonalization is performed separately for each expert even when several of the same shape are batched into one GEMM.

\textbf{Owner assignment.} Of the 4538 Muon matrices of the whole model, 4320 are expert matrices. There are two ways for expert matrices to become assignment units: all 60 experts of one fused tensor form a single unit sharing one owner, or every expert matrix forms its own unit. \mbox{DMuon-2} handles expert matrices separately and takes the latter, so a layer's NS and parameter publish are spread over several ranks. On 256 GPUs the ratio of the busiest owner's share to the mean is 3.757 and 1.078 under the two schemes, and the owner window per rank is 1022.3 MiB and 110.0 MiB. Once the experts are separated, a layer's publish goes from three segments to 180; the operators of Section 5 fold all segments of a layer into one kernel group, so the number of communications is the same as when they are merged.

\section{Experimental Details and Full Results}

This appendix gives the configuration, measurement method, and full results of every experiment in Section 7. All figures in Section 7 are drawn from these tables. D.1 to D.3 cover the configuration and the measurement method, D.4 to D.8 give the full numbers behind every experiment of Section 7, and D.9 and D.10 are two additions beyond the body: the comparison with the concurrent MatrixFSDP, and the MoE model under pure data parallelism.

\subsection{Platform and Models}

The platform has up to 32 nodes, each with eight H100 80GB GPUs connected by NVLink and one 400 Gb/s RDMA NIC per GPU. Cross-node RDMA is initiated by the GPU, which also writes the NIC doorbell directly (IBGDA). Table~\ref{tab:1} lists the training configuration of the six models. In the throughput experiments the LLMs are trained on random tokens. The VLA models use a fixed batch of LIBERO~\citep{liu2023libero} samples prepared in advance, bypassing the data loader, so step times exclude data loading. Qwen2.5-7B uses a sequence length of 2048 under HSDP-Z2, because at 2560 the memory of A exceeds the GPU's capacity; Qwen1.5-MoE-A2.7B does the same under HSDP-Z3. The parameters of Qwen2.5-32B do not fit on one GPU unless sharded, so it runs only under FSDP-Z3 and HSDP-Z3.

\begin{table*}[t]
\centering
\caption{Training configuration of each model.\label{tab:1}}
\small
\begin{tabularx}{\textwidth}{lcYcY}
\toprule
\textbf{Model} & \textbf{Type} & \textbf{Sequence length or input} & \textbf{Micro-batch per GPU} & \textbf{Sharding} \\
\midrule
Qwen2.5-1.5B & LLM & 4096 & 6 & FSDP-Z2, FSDP-Z3, HSDP-Z2 \\
Qwen2.5-7B & LLM & 2560 (2048 under HSDP-Z2) & 4 & FSDP-Z2, FSDP-Z3, HSDP-Z2 \\
Qwen2.5-32B & LLM & 4096 & 4 & FSDP-Z3, HSDP-Z3 \\
Wall-OSS-0.5 & VLA & two 224$\times$224 images, 224 tokens & 36 & FSDP-Z2, FSDP-Z3, HSDP-Z2 \\
$\pi_{0.5}$ & VLA & two 224$\times$224 images, 200 text tokens & 32 & FSDP-Z2, FSDP-Z3, HSDP-Z2 \\
Qwen1.5-MoE-A2.7B & MoE LLM & 2560, 2048 under HSDP-Z3 & 4 & FSDP-Z3, HSDP-Z3 \\
\bottomrule
\end{tabularx}
\end{table*}

\subsection{Implementation of the Four Lines}

A uses PyTorch FSDP2, calling \texttt{fully\_shard} on every layer, with AdamW as the optimizer and the sharding level set by FSDP's sharding strategy. In B, C, and D the \mbox{DMuon-2} runtime takes over the sharding and communication of all trainable parameters and never calls \texttt{fully\_shard}; the three lines share the same code and the same build.

B follows the first route of the introduction. Before the optimizer step it gathers the gradient and momentum of each matrix from all ranks into a whole matrix, runs the full NS on every GPU, and writes the update back to the shard it is responsible for. Compared with C it performs one more gather and one more write-back, and it repeats NS on every GPU. By design B needs only a constant number of temporary buffers the size of the largest matrix beyond C. The memory figures in Table~\ref{tab:2} are nvidia-smi peaks, which include memory reserved by the caching allocator, so the difference between B and C is positive in some configurations and negative in others.

C uses the matrix owners of Sections 3 and 4 and communicates through NCCL. When gathering parameters, it groups a layer's broadcasts into one call inside a node and issues one broadcast per owner across nodes, the faster pattern at each level in our measurements. HSDP shards only within a node, so its parameter gathering is entirely intra-node and is always grouped. D replaces the parameter gathering, the gradient reduction, and the cross-node publish of C with the three operators of Section 5.

\subsection{Measurement Method}

In the end-to-end experiments, the step time is the median step time within the measurement window, over the 60 steps after warmup for an LLM and over the steps after step 30 for a VLA model. Memory is the maximum of the nvidia-smi readings over all GPUs during the whole run, sampled every 2 s and including initialization and warm-up, so spikes shorter than the interval may be missed. On 256 GPUs some runs show bursts of slow steps, including A and C, which do not use our operators, and the median keeps these bursts from deciding a comparison. Each configuration runs once, and we keep only runs that exit normally; the 30 configurations run two to four times report the median of their runs, which differ by 0.3\% at the median and by at most 3.7\% except for one A configuration at 10\% and one B configuration at 9\%. NS kernel times are collected over three additional steps after the measurement window, timed per phase with CUDA events, and we report the slowest rank.

Before each timed point of a communication operator, the GPU stream first waits 2 ms in every column so that the host can enqueue all calls under test, which keeps host-side launch gaps out of the measurement. Each cell runs 20 warm-up and 30 timed iterations and takes the median over iterations of the slowest rank's time; each cell is launched 5 times independently and the median of the launches is reported. Each run also checks the result: the gathered parameters equal the owner's data bit for bit, and the reduction of integer inputs equals the exact sum bit for bit. On 256 GPUs the integer partial sums along the chain exceed the range that BF16 represents exactly, so the reduction is instead checked separately with all-one and all-two inputs, whose partial sums are exact in any order of addition. Column headers give the total bytes of one collective. The variable-length shape places one owner on each node and cuts the message into segments whose lengths cycle in the ratio 1:2:3. These are operator-level microbenchmarks and do not translate directly into end-to-end gains.

The numerical-equivalence experiments compare the loss of D and B step by step. They comprise a long-horizon comparison over 10000 steps on two models (D.8) and a short-horizon check of 60 steps on random tokens in 13 configurations. Reduction error is measured against the float64 exact sum.

\subsection{End-to-End Results}

Table~\ref{tab:2} gives the step time and memory of every configuration; the statistics in Section 7.2 are computed from the rows not marked *. For memory, the only extra cost of D over C is the symmetric pool, 1.0 to 7.5 GiB with a median of about 2.0 GiB, and Qwen2.5-32B under HSDP-Z3 does not fit with A or B, so only C and D run.

The memory of D is well balanced across GPUs (Table~\ref{tab:2}(c)). Over all configurations the heaviest GPU exceeds the per-GPU mean by at most 2.4\%, with a median of 0.3\%, and the heaviest and lightest GPUs differ by at most 2.5 GiB, the least balanced being Qwen2.5-32B. Assigning whole matrices to owners does not make any GPU noticeably heavier.

The two rows marked * in Table~\ref{tab:2} are an MoE model, excluded from the statistics above; its results are in D.10.

\begin{table*}[t]
\centering
\caption{End-to-end step time and memory, behind Figure~\ref{fig:6}. The four numbers in each cell of (a) and (b) are A/B/C/D, (c) lists D only, and a dash marks a configuration that does not fit in memory. In cells marked $\ddagger$ the memory of A is close to the GPU limit, and Section 7 excludes them from its statistics; in cells marked \S{} the memory of D is close to the limit. The B value marked \P{} captured only the steady state: on the other GPU counts the same line shows a brief transient of 76.8 to 77.3 GiB that the sampling missed here, and its steady-state reading of 69.0 GiB matches the 69.0 to 69.5 GiB of the others. The two rows marked * are an MoE model, whose number of Muon matrices is of a different order from the first five models; Section 7 and Figure~\ref{fig:6} exclude both rows from their statistics, and D.10 discusses the results separately. The two rows use different sequence lengths (Table~\ref{tab:1}) and are not compared with each other.\label{tab:2}}
\scriptsize
\setlength{\tabcolsep}{4pt}
\begin{subtable}{\textwidth}
\centering
\caption{Time of one step (ms)}
\begin{tabular*}{\textwidth}{@{\extracolsep{\fill}}lcccccc}
\toprule
\textbf{Model} & \textbf{Sharding} & \textbf{16 GPUs} & \textbf{32 GPUs} & \textbf{64 GPUs} & \textbf{128 GPUs} & \textbf{256 GPUs} \\
\midrule
Qwen2.5-1.5B & FSDP-Z2 & 490/692/509/494 & 490/743/506/491 & 503/814/514/492 & 506/950/529/496 & 527/1265/592/511 \\
Qwen2.5-1.5B & FSDP-Z3 & 497/714/532/504 & 501/773/538/507 & 505/930/566/504 & 523/1037/611/513 & 554/1430/747/515 \\
Qwen2.5-1.5B & HSDP-Z2 & 490/703/516/501 & 491/735/507/500 & 498/807/509/503 & 524/949/512/514 & 522/1230/526/520 \\
Qwen2.5-7B & FSDP-Z2 & 818/1662/863/825 & 816/1692/849/808 & 821/1779/858/800 & 837/1963/887/800 & 865/2346/949/812 \\
Qwen2.5-7B & FSDP-Z3 & 851/1702/904/846 & 850/1749/896/834 & 850/1833/915/823 & 873/2050/976/823 & 900/2492/1094/851 \\
Qwen2.5-7B & HSDP-Z2 & 686$\ddagger$/1691/707/690 & 690$\ddagger$/1721/690/686 & 697$\ddagger$/1809/701/681 & 706$\ddagger$/1999/706/685 & 716$\ddagger$/2401/765/704 \\
Qwen2.5-32B & FSDP-Z3 & 7230/12115/7621/7369 & 7187/12207/7552/7263 & 7235/12438/7549/7156 & 7260/12927/7770/7138 & 7345/13854/8023/7167 \\
Qwen2.5-32B & HSDP-Z3 & ---/---/7629/7501 & ---/---/7528/7479 & ---/---/7480/7482 & ---/---/7500/7466 & ---/---/7574/7352 \\
Wall-OSS-0.5 & FSDP-Z2 & 451/936/467/439 & 457/1024/467/430 & 467/1197/490/428 & 479/1539/543/428 & 523/2146/592/430 \\
Wall-OSS-0.5 & FSDP-Z3 & 473/963/494/450 & 480/1061/501/442 & 496/1260/542/437 & 513/1652/662/444 & 596/2345/792/444 \\
Wall-OSS-0.5 & HSDP-Z2 & 457/949/452/442 & 460/1026/442/435 & 464/1180/440/430 & 478/1475/443/433 & 494/2060/450/440 \\
$\pi_{0.5}$ & FSDP-Z2 & 1013/1338/1027/1004 & 1016/1391/1029/999 & 1016/1507/1046/1000 & 1022/1740/1070/1003 & 1040/2178/1130/1011 \\
$\pi_{0.5}$ & FSDP-Z3 & 1032/1362/1060/1014 & 1034/1433/1067/1005 & 1035/1565/1097/1008 & 1045/1809/1140/1013 & 1068/2273/1229/1020 \\
$\pi_{0.5}$ & HSDP-Z2 & 1014/1340/1011/1011 & 1016/1393/1010/1010 & 1019/1499/1009/1011 & 1021/1714/1017/1018 & 1027/2114/1032/1016 \\
Qwen1.5-MoE-A2.7B* & FSDP-Z3 & 633/4121/978/885 & 628/4894/930/754 & 631/6606/1050/726 & 948/9922/1058/713 & 1602/16528/1211/740 \\
Qwen1.5-MoE-A2.7B* & HSDP-Z3 & 551/3893/733/717 & 568/4965/679/649 & 573/6601/675/630 & 574/9804/641/590 & 582/16401/700/593 \\
\bottomrule
\end{tabular*}
\end{subtable}
\vspace{6pt}
\begin{subtable}{\textwidth}
\centering
\caption{Peak memory per GPU (GiB)}
\begin{tabular*}{\textwidth}{@{\extracolsep{\fill}}lcccccc}
\toprule
\textbf{Model} & \textbf{Sharding} & \textbf{16 GPUs} & \textbf{32 GPUs} & \textbf{64 GPUs} & \textbf{128 GPUs} & \textbf{256 GPUs} \\
\midrule
Qwen2.5-1.5B & FSDP-Z2 & 69.1/68.6/68.8/69.8 & 68.7/68.3/68.4/69.4 & 68.6/68.5/68.2/69.5 & 68.4/68.4/68.2/69.7 & 68.3/68.4/68.2/70.6 \\
Qwen2.5-1.5B & FSDP-Z3 & 66.6/66.3/66.4/67.6 & 66.2/66.0/66.1/67.2 & 66.1/65.8/65.9/67.3 & 65.9/66.1/65.8/67.6 & 65.8/66.1/65.8/68.5 \\
Qwen2.5-1.5B & HSDP-Z2 & 70.8/77.3/69.5/70.5 & 70.8/76.8/69.3/70.4 & 70.8/69.1\P{}/69.2/70.5 & 70.8/76.8/69.1/70.8 & 70.9/76.8/69.1/71.5 \\
Qwen2.5-7B & FSDP-Z2 & 76.4/71.9/72.7/74.7 & 74.3/69.9/70.4/72.3 & 73.4/69.5/69.7/71.8 & 72.9/69.5/69.3/71.7 & 72.6/69.5/69.3/72.6 \\
Qwen2.5-7B & FSDP-Z3 & 64.5/60.4/61.2/63.9 & 62.5/58.4/58.9/61.5 & 61.6/58.0/58.2/61.0 & 61.1/58.0/57.8/60.8 & 60.8/58.0/57.8/61.7 \\
Qwen2.5-7B & HSDP-Z2 & 78.4/74.8/65.6/68.0 & 78.3/73.2/63.6/65.6 & 78.3/72.5/63.0/65.2 & 79.1/72.6/62.6/65.2 & 78.3/72.6/62.6/65.9 \\
Qwen2.5-32B & FSDP-Z3 & 76.5/64.0/66.3/72.9 & 62.2/56.0/57.0/63.2 & 58.0/53.0/53.7/59.0 & 56.3/52.2/52.5/57.3 & 55.2/51.7/51.7/57.2 \\
Qwen2.5-32B & HSDP-Z3 & ---/---/76.2/79.2\S{} & ---/---/67.2/74.2 & ---/---/64.1/70.7 & ---/---/62.8/70.3 & ---/---/62.0/69.5 \\
Wall-OSS-0.5 & FSDP-Z2 & 65.4/63.5/63.6/64.8 & 63.2/62.3/62.3/63.4 & 62.2/61.8/61.6/63.0 & 61.9/61.6/61.4/63.1 & 61.7/61.6/61.4/64.0 \\
Wall-OSS-0.5 & FSDP-Z3 & 57.7/56.5/56.7/58.0 & 55.7/55.3/55.3/56.6 & 54.7/54.8/54.6/56.1 & 54.2/54.6/54.4/56.2 & 54.0/54.6/54.4/57.1 \\
Wall-OSS-0.5 & HSDP-Z2 & 70.9/65.1/65.4/66.4 & 71.0/64.2/64.2/65.3 & 71.1/63.8/63.6/64.9 & 71.1/63.6/63.5/65.0 & 71.1/63.6/63.5/65.7 \\
$\pi_{0.5}$ & FSDP-Z2 & 73.0/71.7/72.0/73.9 & 71.2/71.1/70.3/72.2 & 70.6/70.6/70.1/72.2 & 69.7/69.9/70.1/72.5 & 69.5/70.6/70.0/73.3 \\
$\pi_{0.5}$ & FSDP-Z3 & 68.0/67.2/67.5/69.4 & 65.8/66.6/65.8/67.7 & 65.0/66.1/65.7/67.7 & 64.6/65.5/65.6/68.0 & 64.5/66.1/65.5/68.8 \\
$\pi_{0.5}$ & HSDP-Z2 & 76.9/73.2/73.6/75.2 & 76.9/72.9/72.9/74.8 & 76.9/72.7/72.6/74.6 & 76.9/72.7/72.6/74.9 & 77.1/72.7/72.6/75.6 \\
Qwen1.5-MoE-A2.7B* & FSDP-Z3 & 64.6/58.5/57.0/64.5 & 56.1/54.2/53.5/60.4 & 54.1/54.3/51.5/58.3 & 58.0/54.2/51.1/58.0 & 67.3/54.2/50.5/58.2 \\
Qwen1.5-MoE-A2.7B* & HSDP-Z3 & 73.9/56.5/56.6/68.5 & 73.9/54.8/54.8/66.0 & 73.9/53.8/53.7/64.7 & 73.9/53.4/53.2/64.3 & 73.9/53.1/53.0/64.4 \\
\bottomrule
\end{tabular*}
\end{subtable}
\vspace{6pt}
\begin{subtable}{\textwidth}
\centering
\caption{Per-GPU peak-memory balance of D (GiB): heaviest GPU / mean over GPUs, with the heaviest GPU's excess over the mean in parentheses}
\begin{tabularx}{\textwidth}{lcYYYYY}
\toprule
\textbf{Model} & \textbf{Sharding} & \textbf{16 GPUs} & \textbf{32 GPUs} & \textbf{64 GPUs} & \textbf{128 GPUs} & \textbf{256 GPUs} \\
\midrule
Qwen2.5-1.5B & FSDP-Z2 & 69.8 / 69.7 (+0.1\%) & 69.4 / 69.3 (+0.1\%) & 69.5 / 69.4 (+0.1\%) & 69.7 / 69.7 (+0.1\%) & 70.6 / 70.6 (+0.1\%) \\
Qwen2.5-1.5B & FSDP-Z3 & 67.6 / 67.5 (+0.1\%) & 67.2 / 67.2 (+0.1\%) & 67.3 / 67.2 (+0.1\%) & 67.6 / 67.5 (+0.1\%) & 68.5 / 68.4 (+0.1\%) \\
Qwen2.5-1.5B & HSDP-Z2 & 70.5 / 70.5 (+0.1\%) & 70.4 / 70.3 (+0.2\%) & 70.5 / 70.3 (+0.3\%) & 70.8 / 70.6 (+0.2\%) & 71.5 / 71.3 (+0.3\%) \\
Qwen2.5-7B & FSDP-Z2 & 74.7 / 74.3 (+0.5\%) & 72.3 / 72.2 (+0.2\%) & 71.8 / 71.6 (+0.3\%) & 71.7 / 71.5 (+0.2\%) & 72.6 / 72.3 (+0.3\%) \\
Qwen2.5-7B & FSDP-Z3 & 63.9 / 63.5 (+0.6\%) & 61.5 / 61.3 (+0.3\%) & 61.0 / 60.8 (+0.3\%) & 60.8 / 60.7 (+0.2\%) & 61.7 / 61.5 (+0.4\%) \\
Qwen2.5-7B & HSDP-Z2 & 68.0 / 67.3 (+1.0\%) & 65.6 / 65.5 (+0.3\%) & 65.2 / 65.0 (+0.3\%) & 65.2 / 64.9 (+0.4\%) & 65.9 / 65.6 (+0.5\%) \\
Qwen2.5-32B & FSDP-Z3 & 72.9 / 72.2 (+1.0\%) & 63.2 / 61.7 (+2.4\%) & 59.0 / 58.2 (+1.3\%) & 57.3 / 57.0 (+0.7\%) & 57.2 / 56.9 (+0.4\%) \\
Qwen2.5-32B & HSDP-Z3 & 79.2 / 78.7 (+0.5\%) & 74.2 / 73.2 (+1.3\%) & 70.7 / 70.0 (+1.0\%) & 70.3 / 68.7 (+2.3\%) & 69.5 / 68.0 (+2.2\%) \\
Wall-OSS-0.5 & FSDP-Z2 & 64.8 / 64.6 (+0.2\%) & 63.4 / 63.3 (+0.2\%) & 63.0 / 62.9 (+0.2\%) & 63.1 / 62.9 (+0.3\%) & 64.0 / 63.6 (+0.5\%) \\
Wall-OSS-0.5 & FSDP-Z3 & 58.0 / 57.8 (+0.4\%) & 56.6 / 56.5 (+0.2\%) & 56.1 / 56.0 (+0.2\%) & 56.2 / 56.0 (+0.3\%) & 57.1 / 56.8 (+0.5\%) \\
Wall-OSS-0.5 & HSDP-Z2 & 66.4 / 66.3 (+0.1\%) & 65.3 / 65.1 (+0.3\%) & 64.9 / 64.7 (+0.2\%) & 65.0 / 64.8 (+0.3\%) & 65.7 / 65.3 (+0.5\%) \\
$\pi_{0.5}$ & FSDP-Z2 & 73.9 / 73.0 (+1.2\%) & 72.2 / 72.1 (+0.1\%) & 72.2 / 71.7 (+0.7\%) & 72.5 / 71.5 (+1.3\%) & 73.3 / 72.3 (+1.4\%) \\
$\pi_{0.5}$ & FSDP-Z3 & 69.4 / 68.5 (+1.2\%) & 67.7 / 67.6 (+0.1\%) & 67.7 / 67.2 (+0.7\%) & 68.0 / 67.1 (+1.4\%) & 68.8 / 67.8 (+1.5\%) \\
$\pi_{0.5}$ & HSDP-Z2 & 75.2 / 74.9 (+0.4\%) & 74.8 / 73.8 (+1.2\%) & 74.6 / 73.6 (+1.3\%) & 74.9 / 73.8 (+1.4\%) & 75.6 / 74.4 (+1.6\%) \\
Qwen1.5-MoE-A2.7B* & FSDP-Z3 & 64.5 / 64.5 (+0.0\%) & 60.4 / 60.3 (+0.1\%) & 58.3 / 58.1 (+0.3\%) & 58.0 / 57.9 (+0.2\%) & 58.2 / 58.1 (+0.2\%) \\
Qwen1.5-MoE-A2.7B* & HSDP-Z3 & 68.5 / 68.4 (+0.2\%) & 66.0 / 65.9 (+0.2\%) & 64.7 / 64.6 (+0.2\%) & 64.3 / 64.2 (+0.1\%) & 64.4 / 64.4 (+0.1\%) \\
\bottomrule
\end{tabularx}
\end{subtable}
\end{table*}

\subsection{NS Time and Optimizer-State Memory}

Table~\ref{tab:3} reports how these two quantities change with the GPU count for Qwen2.5-1.5B under HSDP-Z2, behind the discussion in Section 4. The node-local row of Table~\ref{tab:3} has no corresponding runtime switch. We fit a per-shape cost by least squares from the measured NS time and the matrices held by each rank in the same runs, and then compute the slowest rank when the matrices are assigned among the 8 ranks of a node. Applied to the measured cluster-wide assignment, this fit underestimates by 9\% to 21\%, so the estimate is a lower bound.

\begin{table*}[t]
\centering
\caption{NS kernel time and optimizer-state memory per GPU for Qwen2.5-1.5B under HSDP-Z2, behind Figure~\ref{fig:7}(a)(b). NS kernel time is that of the slowest rank. The cluster-wide state memory is measured on the busiest rank; the per-node figure is computed from the placement of plain HSDP, covering one copy of the Muon momentum per node and the AdamW state of layers such as the embedding.\label{tab:3}}
\small
\begin{tabular}{lccccc}
\toprule
\textbf{Quantity} & \textbf{16 GPUs} & \textbf{32 GPUs} & \textbf{64 GPUs} & \textbf{128 GPUs} & \textbf{256 GPUs} \\
\midrule
NS kernel time, cluster-wide owners (ms) & 10.7 & 6.2 & 4.5 & 2.6 & 2.7 \\
NS kernel time, node-local owners (ms, estimate) & 18.8 & 18.8 & 18.8 & 18.8 & 18.8 \\
Optimizer state, one copy per cluster (GiB, busiest rank) & 0.546 & 0.392 & 0.320 & 0.269 & 0.269 \\
Optimizer state, one copy per node (GiB, computed) & 0.813 & 0.813 & 0.813 & 0.813 & 0.813 \\
\bottomrule
\end{tabular}
\end{table*}

\subsection{Step-Time Breakdown}

Table~\ref{tab:4} splits a step into six segments, taken from the timeline of rank 0. Each compute kernel is attributed to the forward pass, the backward pass, or optimizer computation according to the phase that launched it. Exposed communication is communication time hidden behind no computation and not counted as publish wait, and its gradient-reduction and cross-node-publish parts are listed separately. Publish wait is the time the forward pass waits before it starts for the previous step's publish to finish, and idle is time with no kernel running on the GPU.

\begin{figure}[t]
\centering
\includegraphics[width=\columnwidth]{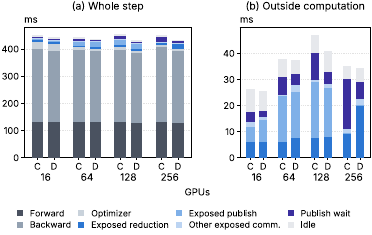}
\caption{Breakdown of one step of Wall-OSS-0.5 under HSDP-Z2, with two bars per GPU count for C and D. (a) The whole step, split into forward, backward, optimizer computation, exposed communication, publish wait, and idle, with exposed communication split into gradient reduction, cross-node publish, and the rest. (b) The part outside computation, enlarged. Values are in Appendix Table~\ref{tab:4}.\label{fig:10}}
\end{figure}

Figure~\ref{fig:10} splits a step into six segments. C and D spend nearly the same time in the forward pass, the backward pass, and optimizer computation, and they differ in communication. On 128 GPUs D cuts the publish wait from 10.2 ms under C to 4.8 ms. This time used to sit exposed at the start of the forward pass, and the persistent kernel overlaps it with the previous step. D also cuts the exposed cross-node publish from 21.4 ms to 18.6 ms, while the exposed gradient reduction is about the same in the two, 7.7 ms against 8.0 ms, so D's step is about 3\% shorter. On 256 GPUs nearly all of C's cross-node publish shows up as publish wait, 19.3 ms, against 6.3 ms for D. D is 6.7 ms faster in the forward pass and 7.9 ms faster in the backward pass, its exposed gradient reduction is 11.0 ms longer (20.2 against 9.2 ms), and the step times are 435.3 against 452.2 ms. Giving the pool reduction fewer thread blocks eases the slowdown but makes the reduction itself slower, and with 8 blocks the step becomes 46\% longer, so 16 blocks is the better trade-off. The step time of B is dominated by the exposed cross-node publish, which takes half of it on 128 GPUs.

\begin{table*}[t]
\centering
\caption{Breakdown of one step of Wall-OSS-0.5 under HSDP-Z2 (ms), behind Figure~\ref{fig:10} and Figure~\ref{fig:7}(c). The share of each segment is the per-step median over the profiling window (9 steps on 16 and 64 GPUs, 7 on 128 and 256), rescaled to the median step time of the same run outside the window to remove the overhead of the profiler. B on 16 GPUs is especially sensitive to profiler overhead, so it is captured with the lighter nsys and rescaled to the step time of a run without capture at the same time. Step times differ from Table~\ref{tab:2} by at most 3.3\%, except for A on 256 GPUs, measured at a different time, which is 4.8\% higher. The C and D rows were captured again under the platform configuration described in D.1, and the A and B rows are from the earlier capture. A has no cross-node publish.\label{tab:4}}
\scriptsize
\begin{tabular}{lcccccccccc}
\toprule
\textbf{GPUs} & \textbf{Line} & \textbf{Forward} & \textbf{Backward} & \textbf{Optimizer} & \textbf{Exposed comm.} & \textbf{of which reduction} & \textbf{of which publish} & \textbf{Publish wait} & \textbf{Idle} & \textbf{Step} \\
\midrule
16 & A & 139.7 & 285.5 & 6.5 & 14.4 & 8.9 & --- & --- & 12.7 & 458.8 \\
16 & B & 128.2 & 260.0 & 296.7 & 207.7 & 7.2 & 193.9 & 0.9 & 59.8 & 953.3 \\
16 & C & 131.2 & 267.5 & 26.9 & 13.8 & 6.2 & 5.5 & 3.8 & 8.8 & 454.8 \\
16 & D & 131.0 & 261.5 & 25.6 & 15.6 & 6.2 & 8.2 & 2.5 & 7.5 & 444.6 \\
64 & A & 138.9 & 291.6 & 6.5 & 18.6 & 10.9 & --- & --- & 17.2 & 472.7 \\
64 & B & 129.1 & 267.5 & 298.7 & 442.6 & 10.8 & 417.2 & 1.2 & 38.0 & 1177.1 \\
64 & C & 131.0 & 265.7 & 9.4 & 24.0 & 5.9 & 18.0 & 6.9 & 6.8 & 445.8 \\
64 & D & 130.5 & 262.6 & 7.8 & 27.9 & 7.4 & 17.7 & 4.2 & 5.4 & 437.7 \\
128 & A & 134.1 & 299.6 & 6.3 & 30.2 & 19.3 & --- & --- & 21.3 & 491.5 \\
128 & B & 127.3 & 266.5 & 296.6 & 745.8 & 12.6 & 705.0 & 1.0 & 46.4 & 1483.6 \\
128 & C & 130.8 & 266.9 & 8.9 & 29.9 & 7.7 & 21.4 & 10.2 & 6.8 & 457.6 \\
128 & D & 126.5 & 258.2 & 8.2 & 28.2 & 8.0 & 18.6 & 4.8 & 8.0 & 442.8 \\
256 & A & 128.5 & 290.9 & 6.2 & 64.6 & 46.1 & --- & --- & 27.5 & 517.7 \\
256 & B & 127.1 & 268.5 & 295.2 & 1354.3 & 14.2 & 1286.5 & 0.8 & 49.5 & 2095.4 \\
256 & C & 133.1 & 272.5 & 9.1 & 10.9 & 9.2 & 0.1 & 19.3 & 4.8 & 452.2 \\
256 & D & 126.4 & 264.6 & 9.0 & 22.6 & 20.2 & 0.1 & 6.3 & 5.6 & 435.3 \\
\bottomrule
\end{tabular}
\end{table*}

\subsection{Communication Operators}

Each cell of Tables 5 and 6 holds four numbers in two rows. The upper row compares two implementations of the same semantics, one collective per owner on the left and our operator on the right. The lower row gives standard primitives as a reference, the single-owner primitive on the left and the equal-length multi-owner primitive on the right. The two primitives in the lower row require equal lengths and a fixed layout, so their semantics differ from the variable-length case, and they only mark what the standard library reaches on the most regular shapes. When the upper-right entry is slower than the lower row at some size, a regular layout is faster at that size. We use per-owner calls as the baseline rather than a coalesced call, because the two differ by a few percent at most with a single owner, while on large variable-length messages a coalesced call becomes much slower.

The loads in Table~\ref{tab:7} come from one rail of each of the five models in HSDP training (the rail of rank 0 on 8 nodes), replaying all publish tasks of a step as they occur in training; each subtable title gives the parameter count, the total bytes, and the share of the largest parameter. Only the node count changes, and the chain depth is the node count minus one.

\begin{table*}[t]
\centering
\caption{The variable-length operator AllGatherV in the forward broadcast direction (ms), behind Figure~\ref{fig:8}(a)(b). In each cell the upper line is per-owner / ours and the lower line is \texttt{ncclBroadcast} / \texttt{ncclAllGather}.\label{tab:5}}
\footnotesize
\setlength{\tabcolsep}{4pt}
\begin{subtable}{\textwidth}
\centering
\caption{Small messages}
\begin{tabularx}{\textwidth}{lYYYY}
\toprule
\textbf{GPUs} & \textbf{16 KiB} & \textbf{64 KiB} & \textbf{256 KiB} & \textbf{1 MiB} \\
\midrule
16 & 0.078 / \textbf{0.033}\newline 0.041 / 0.051 & 0.100 / \textbf{0.034}\newline 0.048 / 0.059 & 0.119 / \textbf{0.036}\newline 0.063 / 0.071 & 0.130 / \textbf{0.041}\newline 0.080 / 0.136 \\
\addlinespace[3pt]
64 & 0.395 / \textbf{0.055}\newline 0.183 / 0.242 & 0.474 / \textbf{0.057}\newline 0.235 / 0.249 & 0.704 / \textbf{0.059}\newline 0.230 / 0.263 & 0.859 / \textbf{0.063}\newline 0.271 / 0.352 \\
\addlinespace[3pt]
128 & 0.738 / \textbf{0.088}\newline 0.351 / 0.465 & 0.830 / \textbf{0.088}\newline 0.482 / 0.479 & 1.140 / \textbf{0.090}\newline 0.416 / 0.492 & 1.674 / \textbf{0.097}\newline 0.496 / 0.662 \\
\addlinespace[3pt]
256 & 1.632 / \textbf{0.151}\newline 0.725 / 0.945 & 1.625 / \textbf{0.151}\newline 0.970 / 0.951 & 1.968 / \textbf{0.153}\newline 0.956 / 0.970 & 2.946 / \textbf{0.171}\newline 0.946 / 1.048 \\
\bottomrule
\end{tabularx}
\end{subtable}
\vspace{6pt}
\begin{subtable}{\textwidth}
\centering
\caption{Large messages}
\begin{tabularx}{\textwidth}{lYYYY}
\toprule
\textbf{GPUs} & \textbf{16 MiB} & \textbf{64 MiB} & \textbf{256 MiB} & \textbf{1 GiB} \\
\midrule
16 & 0.273 / \textbf{0.094}\newline 0.200 / 0.146 & 0.480 / \textbf{0.275}\newline 0.388 / 0.374 & 1.192 / \textbf{0.887}\newline 1.271 / 0.898 & 4.128 / \textbf{3.326}\newline 3.724 / 3.196 \\
\addlinespace[3pt]
64 & 0.975 / \textbf{0.198}\newline 0.552 / 0.596 & 1.386 / \textbf{0.290}\newline 0.752 / 0.610 & 2.210 / \textbf{0.886}\newline 1.488 / 1.605 & 5.105 / \textbf{3.204}\newline 5.177 / 3.590 \\
\addlinespace[3pt]
128 & 3.180 / \textbf{0.277}\newline 0.999 / 1.098 & 2.944 / \textbf{0.429}\newline 1.207 / 1.084 & 3.432 / \textbf{0.924}\newline 1.957 / 1.311 & 6.488 / \textbf{3.309}\newline 4.880 / 4.238 \\
\addlinespace[3pt]
256 & 6.473 / \textbf{0.339}\newline 1.919 / 2.244 & 11.228 / \textbf{0.505}\newline 2.142 / 2.354 & 9.245 / \textbf{1.074}\newline 2.901 / 2.478 & 9.109 / \textbf{3.521}\newline 5.859 / 6.438 \\
\bottomrule
\end{tabularx}
\end{subtable}
\end{table*}

\begin{table*}[t]
\centering
\caption{The variable-length operator ReduceScatterV in the backward reduction direction (ms), behind Figure~\ref{fig:8}(c)(d). In each cell the upper line is per-owner / ours and the lower line is \texttt{ncclReduce} / \texttt{ncclReduceScatter}.\label{tab:6}}
\footnotesize
\setlength{\tabcolsep}{4pt}
\begin{subtable}{\textwidth}
\centering
\caption{Small messages}
\begin{tabularx}{\textwidth}{lYYYY}
\toprule
\textbf{GPUs} & \textbf{16 KiB} & \textbf{64 KiB} & \textbf{256 KiB} & \textbf{1 MiB} \\
\midrule
16 & 0.078 / \textbf{0.043}\newline 0.058 / 0.074 & 0.108 / \textbf{0.044}\newline 0.071 / 0.077 & 0.129 / \textbf{0.046}\newline 0.074 / 0.096 & 0.157 / \textbf{0.055}\newline 0.087 / 0.154 \\
\addlinespace[3pt]
64 & 0.378 / \textbf{0.055}\newline 0.193 / 0.239 & 0.450 / \textbf{0.057}\newline 0.244 / 0.246 & 0.678 / \textbf{0.060}\newline 0.227 / 0.263 & 0.832 / \textbf{0.064}\newline 0.258 / 0.353 \\
\addlinespace[3pt]
128 & 0.768 / \textbf{0.095}\newline 0.378 / 0.467 & 0.854 / \textbf{0.096}\newline 0.517 / 0.485 & 1.179 / \textbf{0.097}\newline 0.432 / 0.498 & 1.708 / \textbf{0.100}\newline 0.504 / 0.670 \\
\addlinespace[3pt]
256 & 1.557 / \textbf{0.175}\newline 0.734 / 0.940 & 1.621 / \textbf{0.176}\newline 0.989 / 0.947 & 1.955 / \textbf{0.179}\newline 0.942 / 0.989 & 2.912 / \textbf{0.185}\newline 0.921 / 1.050 \\
\bottomrule
\end{tabularx}
\end{subtable}
\vspace{6pt}
\begin{subtable}{\textwidth}
\centering
\caption{Large messages}
\begin{tabularx}{\textwidth}{lYYYY}
\toprule
\textbf{GPUs} & \textbf{16 MiB} & \textbf{64 MiB} & \textbf{256 MiB} & \textbf{1 GiB} \\
\midrule
16 & 0.258 / \textbf{0.114}\newline 0.189 / 0.173 & 0.460 / \textbf{0.273}\newline 0.376 / 0.430 & 1.164 / \textbf{0.914}\newline 1.164 / 0.823 & 3.814 / \textbf{3.501}\newline 3.515 / 2.832 \\
\addlinespace[3pt]
64 & 0.915 / \textbf{0.244}\newline 0.497 / 0.572 & 1.259 / \textbf{0.348}\newline 0.695 / 0.664 & 2.024 / \textbf{0.956}\newline 1.409 / 1.569 & 4.802 / \textbf{3.433}\newline 4.784 / 3.155 \\
\addlinespace[3pt]
128 & 3.225 / \textbf{0.431}\newline 0.927 / 1.142 & 2.871 / \textbf{0.542}\newline 1.139 / 1.190 & 3.170 / \textbf{1.036}\newline 1.846 / 1.310 & 5.951 / \textbf{3.475}\newline 4.779 / 4.011 \\
\addlinespace[3pt]
256 & 6.347 / \textbf{0.453}\newline 1.743 / 2.245 & 11.137 / \textbf{0.645}\newline 1.984 / 2.225 & 8.766 / \textbf{1.246}\newline 2.704 / 2.379 & 8.402 / \textbf{3.910}\newline 5.598 / 6.288 \\
\bottomrule
\end{tabularx}
\end{subtable}
\end{table*}

\begin{table*}[t]
\centering
\caption{Time of the cross-node publish on one real rail of each of the five models (ms), behind Figure~\ref{fig:8}(e).\label{tab:7}}
\footnotesize
\setlength{\tabcolsep}{4pt}
\begin{subtable}{\textwidth}
\centering
\caption{Qwen2.5-1.5B (HSDP-Z2): 45 parameters, 278.3 MiB, largest 26.2 MiB (9.4\%)}
\begin{tabular*}{\textwidth}{@{\extracolsep{\fill}}lcccc}
\toprule
\textbf{Nodes} & \textbf{Chain depth} & \textbf{Per-owner} & \textbf{Coalesced} & \textbf{Ours} \\
\midrule
2 & 1 & 7.23 & 6.79 & \textbf{3.95} \\
4 & 3 & 9.23 & 9.87 & \textbf{5.57} \\
8 & 7 & 11.73 & 13.26 & \textbf{7.56} \\
16 & 15 & 14.77 & 24.77 & \textbf{8.58} \\
32 & 31 & 20.18 & 43.74 & \textbf{9.78} \\
\bottomrule
\end{tabular*}
\end{subtable}
\vspace{6pt}
\begin{subtable}{\textwidth}
\centering
\caption{Qwen2.5-7B (HSDP-Z2): 38 parameters, 1592.6 MiB, largest 129.5 MiB (8.1\%)}
\begin{tabular*}{\textwidth}{@{\extracolsep{\fill}}lcccc}
\toprule
\textbf{Nodes} & \textbf{Chain depth} & \textbf{Per-owner} & \textbf{Coalesced} & \textbf{Ours} \\
\midrule
2 & 1 & 36.04 & 31.70 & \textbf{19.09} \\
4 & 3 & 39.07 & 43.79 & \textbf{28.68} \\
8 & 7 & 42.81 & 55.70 & \textbf{33.05} \\
16 & 15 & 46.30 & 96.60 & \textbf{38.92} \\
32 & 31 & 55.80 & 181.34 & \textbf{40.07} \\
\bottomrule
\end{tabular*}
\end{subtable}
\vspace{6pt}
\begin{subtable}{\textwidth}
\centering
\caption{Qwen2.5-32B (HSDP-Z3): 55 parameters, 7430.0 MiB, largest 270.0 MiB (3.6\%)}
\begin{tabular*}{\textwidth}{@{\extracolsep{\fill}}lcccc}
\toprule
\textbf{Nodes} & \textbf{Chain depth} & \textbf{Per-owner} & \textbf{Coalesced} & \textbf{Ours} \\
\midrule
2 & 1 & 165.35 & 136.13 & \textbf{83.81} \\
4 & 3 & 175.89 & 196.61 & \textbf{125.75} \\
8 & 7 & 181.42 & 271.49 & \textbf{147.21} \\
16 & 15 & 190.49 & 492.97 & \textbf{174.12} \\
32 & 31 & 213.01 & 955.97 & \textbf{175.71} \\
\bottomrule
\end{tabular*}
\end{subtable}
\vspace{6pt}
\begin{subtable}{\textwidth}
\centering
\caption{Wall-OSS-0.5 (HSDP-Z2): 75 parameters, 870.3 MiB, largest 86.0 MiB (9.9\%)}
\begin{tabular*}{\textwidth}{@{\extracolsep{\fill}}lcccc}
\toprule
\textbf{Nodes} & \textbf{Chain depth} & \textbf{Per-owner} & \textbf{Coalesced} & \textbf{Ours} \\
\midrule
2 & 1 & 20.66 & 20.53 & \textbf{11.53} \\
4 & 3 & 25.22 & 26.88 & \textbf{18.46} \\
8 & 7 & 29.75 & 39.91 & \textbf{20.64} \\
16 & 15 & 37.65 & 69.26 & \textbf{22.23} \\
32 & 31 & 52.97 & 124.60 & \textbf{23.59} \\
\bottomrule
\end{tabular*}
\end{subtable}
\vspace{6pt}
\begin{subtable}{\textwidth}
\centering
\caption{$\pi_{0.5}$ (HSDP-Z2): 83 parameters, 665.3 MiB, largest 128.0 MiB (19.2\%)}
\begin{tabular*}{\textwidth}{@{\extracolsep{\fill}}lcccc}
\toprule
\textbf{Nodes} & \textbf{Chain depth} & \textbf{Per-owner} & \textbf{Coalesced} & \textbf{Ours} \\
\midrule
2 & 1 & 15.98 & 16.12 & \textbf{9.31} \\
4 & 3 & 18.94 & 19.53 & \textbf{13.76} \\
8 & 7 & 22.24 & 32.99 & \textbf{16.55} \\
16 & 15 & 25.66 & 58.95 & \textbf{20.03} \\
32 & 31 & 34.98 & 114.68 & \textbf{22.52} \\
\bottomrule
\end{tabular*}
\end{subtable}
\end{table*}

\subsection{Numerical Equivalence}

Every timed run of the communication operators checks its result: the gathered parameters are bit-identical to the owner's data, the reduction of integer inputs is bit-identical to the exact sum, and on 256 GPUs the reduction is additionally checked with all-one and all-two inputs. The maximum relative error of the reduction against the float64 exact sum is 0.0096 to 0.0165, against 0.0209 to 0.0451 for NCCL's ring reduction under the same protocol. A further 13 configurations were each run for 60 steps on random tokens, where the maximum loss difference between D and B is 0.0013, no more than 0.07\% in relative terms.

\begin{figure*}[t]
\centering
\includegraphics[width=\textwidth]{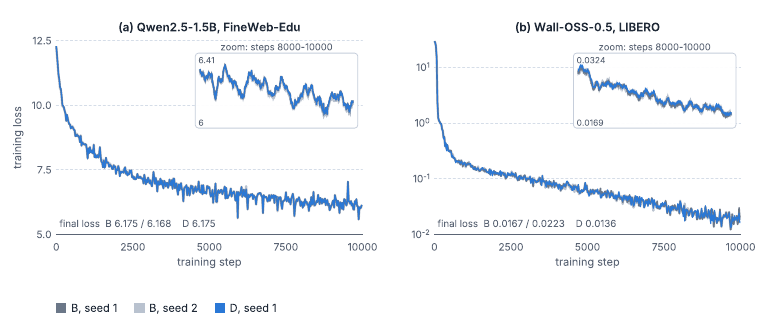}
\caption{Long-horizon convergence over 10000 steps. (a) Qwen2.5-1.5B on FineWeb-Edu, linear vertical axis; (b) Wall-OSS-0.5 on LIBERO, logarithmic vertical axis. Each panel shows three curves, B with two seeds and D; the inset magnifies steps 8000 to 10000 under a 50-step moving average. The three curves coincide over the full range and in the inset: the median per-step relative difference is 0.0028\% (LLM) and 3.12\% (VLA) when the implementation changes, against 0.1022\% and 2.98\% when only the seed changes.\label{fig:11}}
\end{figure*}

Configuration of the long-horizon comparison (Figure~\ref{fig:11}): Qwen2.5-1.5B on FineWeb-Edu and Wall-OSS-0.5 on LIBERO, each with lines B and D for 10000 steps and one node per arm; B was also run with a second seed as a control. Both models use their own upstream training hyperparameters, Wall-OSS-0.5 is initialized from Qwen2.5-VL-3B-Instruct, and the two lines share the same initial weights.

\subsection{Comparison with MatrixFSDP}

MatrixFSDP~\citep{gao2026matrixfsdp}, commit \texttt{2b59736}, places each two-dimensional matrix whole on one owner, runs Muon locally on that owner, and leaves parameter gathering and gradient reduction to off-the-shelf collectives, which is the design point of our C line. Its published experiments cover ZeRO-3 only, so this section measures it under FSDP-Z3 alone and denotes it M.

The model, the data and the measurement share the same code as the D line: the same model construction path, the same fused kernels, FA3, and the same data provider, with the sequence length and per-GPU micro-batch of Table~\ref{tab:1}, gradient accumulation of 1, and full recomputation for Qwen2.5-32B. The optimizer hyperparameters match D, including the Muon learning rate, the $0.2\sqrt{\max(m,n)}$ scaling, Nesterov momentum of 0.95, the Gram NS coefficients and restart, and AdamW for the embedding, the output head, biases and norms. Precision matches D as well, with fp32 parameters and optimizer state and bf16 computation and communication. All of this enters through MatrixFSDP's own extension point for matrix optimizers, leaving its code unchanged. Timing follows D.3.

Everything in the previous paragraph can be matched; three things cannot, all of them features MatrixFSDP does not support, and M pays for each. It does not support shared parameters, so the output head of Qwen2.5-1.5B is untied from the embedding and M gathers and reduces an extra matrix of 233M parameters every step. It does not support meta initialization, so parameters are materialized per unit on the GPU and then sharded, which affects startup only and falls outside the timing window. Its Gram product runs on cuBLAS while D uses a symmetric-product kernel, a difference we previously measured at 0.44\% to 0.75\% end to end.

M's configuration is chosen to give it its fastest option, so that insufficient tuning is not counted against it. The authors' launch scripts provide three presets, two of which use the native collectives of their CUDA extension and were no faster than the baseline preset under matched conditions. We additionally enabled three of their own switches: per-owner broadcasts without coalescing for the parameter gather, owner-segment prefetch, and topology-aware owner placement by node. These three are not an advantage at every scale: Qwen2.5-7B and 32B turn faster from 128 GPUs upward and Qwen2.5-1.5B only at 256, while Wall-OSS-0.5 is faster at every scale. Each cell therefore takes the faster of the two groups, and Table~\ref{tab:8} lists both.

\begin{table*}[t]
\centering
\caption{Step time of MatrixFSDP under FSDP-Z3 (ms). Each cell gives the authors' preset / the tuned configuration, and the value in bold is the one that enters the comparison in the main text. The step times of C and D under the same configuration are the FSDP-Z3 rows of the same four models in Table~\ref{tab:2}.\label{tab:8}}
\small
\begin{tabular}{lccccc}
\toprule
\textbf{Model} & \textbf{16 GPUs} & \textbf{32 GPUs} & \textbf{64 GPUs} & \textbf{128 GPUs} & \textbf{256 GPUs} \\
\midrule
Qwen2.5-1.5B & \textbf{523} / 540 & \textbf{527} / 548 & \textbf{538} / 557 & \textbf{564} / 576 & 635 / \textbf{624} \\
Qwen2.5-7B & \textbf{904} / 937 & \textbf{896} / 930 & \textbf{924} / 944 & 1123 / \textbf{973} & 1760 / \textbf{1025} \\
Qwen2.5-32B & \textbf{7653} / 7766 & \textbf{7548} / 7667 & \textbf{7610} / 7680 & 7822 / \textbf{7723} & 9216 / \textbf{7806} \\
Wall-OSS-0.5 & 560 / \textbf{528} & 710 / \textbf{547} & 1082 / \textbf{599} & 1924 / \textbf{646} & 3530 / \textbf{859} \\
\bottomrule
\end{tabular}
\end{table*}

There is no M on the MoE model of D.10. The owner unit of MatrixFSDP is a two-dimensional parameter, and the predicate of both its planning and its optimizer grouping is a shape that is exactly two-dimensional, while efficient expert computation requires the weights of a layer's experts to be merged into one three-dimensional tensor; such a tensor has no corresponding owner unit there and is routed to AdamW by default. Its released code contains no grouped-GEMM path, its own MoE test executes a set of per-expert two-dimensional linear layers in a loop, and its documentation also suggests leaving routed experts entirely to expert parallelism outside the framework.

\subsection{The MoE Model}

The method of this paper is confined to data parallelism, while MoE training usually also uses expert parallelism. This section reports MoE results under pure data parallelism only, introducing neither expert nor tensor parallelism. The model is Qwen1.5-MoE-A2.7B~\citep{qwen2024moe} with 14.3B parameters in total, 60 experts in each of 24 layers, expert matrices of 1408x2048, a sequence length of 2560 under FSDP-Z3 and 2048 under HSDP-Z3, a per-GPU micro-batch of 4, and a gradient accumulation of 1. How expert weights are handled at the two granularities of computation and optimization is described in C.5. Step times and memory are in the two rows marked * in Table~\ref{tab:2}.

The parameter publish and the gradient reduction of MoE consist of many small communications: the expert weights of one layer total 990 MiB and belong to 180 assignment units (C.5), and issuing one call per owner makes every one of them fill and drain the link again. In one profile taken separately on 256 GPUs, a step of C has 96 Broadcasts and 48 Reductions while D has none of either, and the total number of communication kernels falls from 258 to 162; the per-GPU micro-batch and the activation recomputation of that profile differ from Table~\ref{tab:2}, so it only illustrates the composition of the communication. The gain of D over C grows with scale: under FSDP-Z3 the ratio of step times is 0.905, 0.811, 0.691, 0.674, and 0.611 from 16 to 256 GPUs, and under HSDP-Z3 it is 0.978, 0.956, 0.933, 0.920, and 0.847. Under FSDP-Z3 the step time of C grows by a factor of 1.238 from 16 to 256 GPUs while that of D shrinks by a factor of 0.836, so D gets faster rather than slower with scale over this range. The memory of D exceeds that of C by 6.8 to 7.7 GiB under FSDP-Z3 and by 11.0 to 11.9 GiB under HSDP-Z3, which is the symmetric pool. The comparison with AdamW changes with the form of parallelism: the expert matrices of MoE make the NS computation that Muon adds far larger than on a dense model, so under FSDP-Z3 A is faster up to 64 GPUs; at larger scales the step time of A rises with the GPU count, D is faster from 128 GPUs on, and on 256 GPUs the step time of A is 2.16 times that of D. Under HSDP-Z3 the step time of A barely changes with scale and stays below that of D at all five scales, but the gap narrows from 23\% on 16 GPUs to 1.9\% on 256, at the price of a memory footprint that stays at 73.9 GiB, close to the per-GPU limit.

\end{document}